\RequirePackage{ifpdf}
\ifpdf 
\documentclass[pdftex]{sigma}
\else
\documentclass{sigma}
\fi

\usepackage[all]{xy}

\def\NN{{\mathbb N}}

\def\NN{{\mathbb N}}
\def\ZZ{{\mathbb Z}}
\newcommand{\e}{\varepsilon}
\newcommand{\p}{\varphi}
\newcommand{\ttA}{{\tilde {\tilde A}}}
\newcommand{\tA}{\tilde  A}
\newcommand{\ttB}{{\tilde {\tilde B}}}
\newcommand{\tB}{\tilde  B}
\newcommand{\ttP}{{\tilde {\tilde \varphi}}}
 \newcommand{\tP}{\tilde  \varphi}
\newcommand{\hhA}{{\hat {\hat A}}}
\newcommand{\hA}{\hat  A}
\newcommand{\hhB}{{\hat {\hat B}}}
\newcommand{\hB}{\hat  B}
\newcommand{\hhP}{{\hat {\hat \varphi}}}
 \newcommand{\hP}{\hat  \varphi}
\newcommand{\tpsi}{\tilde \psi}
\newcommand{\ttpsi}{{\tilde{ \tilde \psi}}}
\newcommand{\ttl}{{\tilde{\tilde  \lambda}}}
\newcommand{\tl}{\tilde \lambda}

\newcommand{\g}{{\mathfrak g}}

\newcommand{\3}{3SUSY}

\newcommand{\s}{{{\mathfrak{sl}}(2)}}

\begin{document}
\allowdisplaybreaks

\renewcommand{\PaperNumber}{056}

\FirstPageHeading

\ShortArticleName{Extension of the Poincar\'e Symmetry and Its
Field Theoretical Implementation}

\ArticleName{Extension of the Poincar\'e Symmetry\\ and Its Field
Theoretical Implementation}

\Author{Adrian TANAS\u{A}} \AuthorNameForHeading{A. Tanas\u{a}}

\Address{Laboratoire MIA, Facult\'e de Sciences et Techniques, Universit\'e de Haute Alsace, \\
4 rue des Fr\`eres Lumi\`ere, 68093 Mulhouse Cedex, France}
\Email{\href{mailto:adrian.tanasa@ens-lyon.org}{adrian.tanasa@ens-lyon.org}}

\ArticleDates{Received October 31, 2005, in f\/inal form April 28,
2006; Published online May 29, 2006}

\Abstract{We def\/ine
 a new algebraic extension of the  Poincar\'e symmetry;
 this algebra  is used to implement a f\/ield theoretical model.
 Free Lagrangians are explicitly constructed; several discussions
 regarding degrees of freedom, compatibility with Abelian gauge invariance {\it etc.}
 are done. Finally we analyse the possibilities of interaction terms for this model.}


\Classification{81T60; 17B99}

\section{Introduction and motivation}\label{sec1}

The notion of symmetry has always been playing a key role in
fundamental particle physics. The conventional way to represent
symmetries is, from a mathematical point of view, the notion of
Lie algebra. Indeed, placing oneself within the framework of the
Standard Model and using Lie algebras to group the symmetries of
Nature, one has several powerful no-go theorems. 

One can mention here the O'Raifeartaigh theorem~\cite{o} which 
further led to the Coleman--Mandula theorem~\cite{cm}.
Accepting the assumptions of
the interacting relativistic quantum f\/ield theory
(QFT)\footnote{An interested reader can consult for this purpose
e.g. the book \cite{lopuszanski} of J. Lopuszanski. Note also that
one can  recall here the axiomatic approach to QFT (for example,
the Wightman formalism \cite{uit}).}, the theorem states what are
possible symmetries of Nature. One assumes that the symmetry group
contains the Poincar\'e group as a subgroup, that the $S$-matrix
is based on the local, interacting relativistic quantum f\/ield
theory in $4$ dimensions (thus one has e.g. analytic dependence of
the center-of-mass energy $s$ and invariant momentum transfer $t$
of the elastic-scattering amplitude); that there are only a
f\/inite-number of dif\/ferent particles associated with
one-particle states of  given mass and that there is  energy gap
between  vacuum and  one-particle states. The theorem then states
that  the demanded symmetry group is the direct product of an
internal symmetry group and the Poincar\'e group. These internal
symmetry transformations act only on particle-type indices and
have no  matrix elements between particles of dif\/ferent
four-momentum or dif\/ferent spin.

However, a crucial hypothesis of the Coleman--Mandula theorem
(completely natural at that moment) is  utilisation, at
inf\/initesimal level, of the notion of Lie algebras to group
symmetries. It is exactly on  utilisation of this hypothesis that
we will speculate here.

Independently of all this, it is well-known that  particle physics
today needs to go beyond its Standard Model. One of the most
appealing candidates for New Physics is supersymmetry (SUSY). SUSY
extends the Poincar\'e symmetries by using a larger class of
algebraic structures, Lie superalgebras. The Coleman--Mandula no-go
theorem is therefore not contradicted. Moreover, a somewhat
analogous no-go theorem exists, the Haag--Lopuszanski--Sohnius
theorem \cite{hls} which states that, within the same framework of
interacting relativistic QFT, the only Lie superalgebras extending
the Poincar\'e symmetries are the SUSY extensions. Analogously to
the Coleman--Mandula theorem, we put the emphasis here on  the
fact that this theorem uses Lie superalgebras to group symmetries.

Indeed, one may raise the question of using other type of
algebraic structure to group symmet\-ries. Here we review the use of
a specif\/ic algebraic extension of the Poincar\'e algebra
(introduced in a more formal framework in \cite{Michel-finit,
Michel-infinit}) to the construction of a f\/ield theoretic model.
The model treated here was f\/irst introduced in  by N.~Mohammedi
{\it et.\ al.} in \cite{articol} and further developed by
G.~Moultaka {\it et.\ al.} in~\cite{io1}.

Let us also mention that a dif\/ferent algebraic structure, Lie
parasuperalgebras~\cite{para-alg},
 gave birth to a dif\/ferent f\/ield theoretic model called parasupersymmetry, which was developed by
  J.~Beckers and N.~Debergh in~\cite{paraSUSY}.
In relation to this subject note that parasupersymmetry was
introduced in a non-equivalent way by V.A.~Rubakov and
V.P.~Spiridonov in \cite{paraSUSY-alta}. Furthermore, fractional
supersymmetry was also analysed as a quantum mechanics model by a
series of authors in \cite{FSUSY, FSUSY2, FSUSY4, FSUSY5, FSUSY6,
FSUSY7, FSUSY8, FSUSY9}. Let us recall that a dif\/ferent approach
was given by R.~Kerner in~\cite{Kerner}. Finally let us also
mention that mathematical literature of\/fers another example of
an exotic algebraic structure,  $n$-Lie algebras~\cite{nLie}.

The physical motivation of such an approach is two-sided. First,
the aim would be to develop a self-coherent model which may
of\/fer answers to some of today's open questions of fundamental
physics {\it via} the use of new types of symmetries. However, if
this is proven not to be the case, then such an approach should
lead to  enforcement of the existing no-go theorems.

The attempt presented here shows that there are considerable
dif\/f\/iculties for constructing  such a model. Therefore, as
stated above, one may see this type of approach as a f\/irst step
towards  enforcement of no-go theorems, in the sense that stronger
no-go theorems, which use also other types of algebraic structures
(in this case cubic ones), might be obtained.

This review is organised as follows. In the Section \ref{sec2} we
give the general algebraic setting for this approach. In the
Section~\ref{sec3} several comments are made on the foundation of
this type of exotic approach. Connection with other constructions
are discussed. In the Section \ref{sec4} we construct bosonic
multiplets associated to this structure. In the Section
\ref{4-liber} a free theory is obtained. In the Section
\ref{interactiuni} we study the possibilities of interaction
within the bosonic multiplets obtained previously. Finally, some
technical details are given in the Appendix.

Let us also notice here that this review is a  continuation of
\cite{teza}.

\section{Underlying algebraic structure}\label{sec2}

In this section we give def\/initions for the algebraic setting
used; general aspects are brief\/ly discussed (for more details
one may refer to  \cite{Michel-finit, Michel-infinit} and
\cite{io}).

In  \cite{Michel-finit, Michel-infinit}, a complex Lie algebra of
order $F$ ($F\in\NN^*$) is def\/ined as a ${\mathbb Z}_F$-graded
${\mathbb C}$-vector space ${\mathfrak{g}}= {\mathfrak{g}}_0
\oplus {\mathfrak{g}}_1\oplus {\mathfrak{g}}_2\oplus \dots \oplus
{\mathfrak{g}}_{F-1}$ satisfying the following conditions:
\begin{enumerate}\itemsep=0pt
\item $\mathfrak{g}_0$ is a complex Lie algebra. \item For all
$i\in \{ 1, \dots, F-1 \}$, $\mathfrak{g}_i$ is a representation
of $\mathfrak{g}_0$. \item  For all $i=1,\dots,F-1$ there exists
an $F$-linear,  $\mathfrak{g}_0$-equivariant  map
\[
\mu_i : {\cal S}^F\left(\mathfrak{g}_i\right) \rightarrow
\mathfrak{g}_0,
\]
 where  ${ \cal S}^F(\mathfrak{g}_i)$ denotes
the $F$-fold symmetric product of $\mathfrak{g}_i$, satisfying the
following (Jacobi) identity:
\[
\sum\limits_{j=1}^{F+1} \left[ Y_j,\mu_i ( Y_1,\dots, Y_{j-1},
Y_{j+1},\dots,Y_{F+1}) \right] =0, \nonumber
\]
for any $i=1,\dots,F-1$ and for all $Y_j \in \mathfrak{g}_i$,
$j=1,\ldots,F+1$.
\end{enumerate}

Note that if $F=1$, by def\/inition $\mathfrak{g}=\mathfrak{g}_0$
and a Lie algebra of order $1$ is a Lie algebra. If $F=2$, then
$\g$ is a Lie superalgebra. In this sense, Lie algebras of order
$F$ appear as some kind of generalisations of Lie algebras and
superalgebras.

Note that, by  def\/inition, the following Jacobi identities are
satisf\/ied:
\begin{enumerate}
\itemsep=0pt \item[(i)] For any $ X,X',X'' \in \mathfrak{g}_0$,
\[
\left[\left[X,X'\right],X''\right] +
\left[\left[X',X''\right],X\right] +
\left[\left[X'',X\right],X'\right] =0.
\]
This relation expresses the fact that $\mathfrak{g}_0$ is a Lie
algebra. \item[(ii)] For any $X,X' \in {\mathfrak{g}}_0 \mbox{ and
} Y \in {\mathfrak{g}}_i, i=1,\ldots,F-1$,
\[ \left[\left[X,X'\right],Y\right] +
\left[\left[X',Y\right],X\right] +
\left[\left[Y,X\right],X'\right]=0,
\]
since ${\mathfrak{g}}_i$ is a representation of $\mathfrak{g}_0$.
\item[(iii)] For $X\in {\mathfrak{g}}_0 \mbox{ and } Y_j \in
{\mathfrak{g}}_i$, $j=1,\dots,F$, $i=1,\dots,F-1$
\[
\left[X,\mu_i (Y_1,\dots,Y_F)] = \mu_i (\left[X,Y_1
\right],\dots,Y_F)  + \cdots + \mu_i
(Y_1,\ldots,\left[X,Y_F\right] \right)
\]
(results from  the $\g_0$-equivariance of the map $\mu_i$).
\item[(iv)] For all $Y_j \in \mathfrak{g}_i$, $j=1,\dots,F+1$,
$i=1,\dots,F-1$,
\[
\sum\limits_{j=1}^{F+1} \left[ Y_j,\mu_i ( Y_1,\dots, Y_{j-1},
Y_{j+1},\dots,Y_{F+1}) \right] =0,
\]
which corresponds to condition $3$ of the def\/inition.
\end{enumerate}

Recall here that in the case of Lie (super)algebras Jacobi
identities are included as a necessary element which is related to
the associativity of the corresponding Lie groups. However, in our
case, even though one has corresponding Jacobi identities (see
above), one does not know a~corresponding exponentiation to obtain
some ``group''.

One important thing to notice here is that, if
${\mathfrak{g}}_0\oplus{\mathfrak{g}}_1\oplus\cdots\oplus{\mathfrak{g}}_{F-1}$
is a Lie algebra of order~$F$, with $F>1$, for any
$i=1,\ldots,F-1$, the vector spaces
${\mathfrak{g}}_0\oplus{\mathfrak{g}}_i$ is  also Lie algebra of
order $F$. The particular algebraic extension we use here is of
this type and considers the case $F=3$; we denote the map $\mu$ by
the $3$-entries bracket $\{\cdot,\cdot,\cdot\}$; the algebra
writes:
\begin{gather}
\left[L_{mn}, L_{pq}\right]= i(\eta_{nq} L_{pm}-\eta_{mq} L_{pn} +
\eta_{np}L_{mq}-\eta_{mp} L_{nq}),
\nonumber\\
\left[L_{mn}, P_p \right]= i(\eta_{np} P_m -\eta_{mp} P_n), \nonumber \\
\left[L_{mn}, V_p \right]= i(\eta_{np} V_m -\eta_{mp} V_n), \qquad
\left[P_{m}, V_n \right]= 0, \nonumber\\
\left\{V_m, V_n, V_r \right \}= \eta_{m n} P_r +  \eta_{m r} P_n +
\eta_{r n} P_m,\label{algebra}
\end{gather}
where
\begin{gather}
\label{3-bracket} \{V_m,V_n,V_r \}= V_m V_n V_r + V_m V_r V_n +
V_n V_m V_r + V_n V_r V_m + V_r V_m V_n + V_r V_n V_m
\end{gather}
 stands for the symmetric product of order $3$, $L_{mn}$ are the usual
 generators of the Lorentz algebra ${\mathfrak {so}}(1,3)$, $P_m$ are the momentums ($m,n=0,\dots , 3$) and
$\eta_{mn} = \mathrm{diag}\left(1,-1,-1,-1\right)$  is the
Minkowski metric.

Note that, from the point of view of fundamental physics, there is
no special in this value of $F$; this approach attempts, as was
already stated before, to explore the possibility of constructing
a f\/ield theoretical model based on cubic symmetries.

Comparing to the SUSY extension one can already make a certain
number of comments. As already noticed, by using a dif\/ferent
algebraic structure a construction of this type evades {\it
a~priori} the hypothesis of no-go theorems. Moreover, one can
notice from the second line of~\eqref{algebra} that the
$3$-charges $V$ lie in the vector representation of the Lorentz
algebra and not in the spinor representation, as was the case for
SUSY. This will have as a f\/irst consequence the fact that the
multiplets to be obtained later will be either bosonic or
fermionic (one will not have statistics-mixing multiplets, as is
the case for SUSY).

Another thing to notice  is the following: if the supercharges $Q$
are referred to as  ``square roots of translations'' (since
$Q^2\propto P$, see for example \cite{sohnius}), we can speak of
the generators $V$ as some kind of ``{\it cubic roots of
translations}'' (since $V^3\propto P$).

\section{Comments on the foundation of this approach;\\ connection with other constructions}\label{sec3}

We have thus seen in the previous sections the motivation and the
basis of the foundation of this approach. In this section we argue
more thoroughly on some implications of such a~construction when
reporting it to the usual concepts of QFT.
 We start by having a closer look to the representations of the complexif\/ied Poincar\'e algebra.
 Then, the f\/irst subsection makes a~formal connection between the
generators $V$ and the the supercharges $Q$ of SUSY. The second
subsection refers to the assumption of analytic dependence of a
f\/ield theoretic model.

\subsection{Representations of dimension 4 of the Poincar\'e algebra}

In studying dif\/ferent classif\/ications of the extensions of
Poincar\'e algebra, a key role is played by the study of its
representations. We present here such results for $4$-dimensional
representations, other results also to be found in~\cite{io}.

Notice here that, when considering both our extension and the SUSY
extension\footnote{Obviously, we refer here to the simplest SUSY
extension, $N=1$ and with no central charges (see for
examp\-le~\cite{sohnius}).}, one uses representations of the
Poincar\'e algebra of dimension $4$. We now point out  what are
all these $4$-dimensional representations.

For this, consider the complexif\/ied Poincar\'e algebra. It is
well-known that its semi-simple part is isomorphic to $\s
\oplus\s$. This can be seen explicitly by performing the following
change of basis in the Lorentz algebra:
\begin{gather*}
U_1=iM_{01}-M_{23},  \qquad
U_2=-\frac{1}{2}(M_{02}-M_{12}+iM_{03}-iM_{13}), \\
U_3=\frac{1}{2}(M_{02}+M_{12}-iM_{03}-iM_{13}) ,\\
W_1=iM_{01}+M_{23} ,  \qquad
W_2=-\frac{1}{2}(M_{02}-M_{12}-iM_{03}+iM_{13}) , \\
W_3=\frac{1}{2}(M_{02}+M_{12}+iM_{03}+iM_{13}).
\end{gather*}
The semi-simple part writes
\begin{alignat}{3}
&\left[U_1,U_2\right]=-2U_2,\qquad &&  \left[W_1,W_2\right]=-2W_2, &\nonumber\\
& \left[U_1,U_3\right]=2U_3, &&  \left[W_1,W_3\right]=2W_3,& \nonumber\\
& \left[U_3,U_2\right]=U_1 , && \left[W_3,W_2\right]=W_1.&
\label{2sl2}
\end{alignat}
The basis \eqref{2sl2} is more handy when one  confront  the
problem of writing down all the representations of the
complexif\/ied Poincar\'e algebra. (Indeed, then one can make
direct use of the highest weight representations of~$\s$.)

Let us denote now  the spin $j$ representations of $\s$ by ${\cal
D}_j$. Furthermore, denote by $ {\cal D}_{(j_1,j_2)} $
 the representation of $\s \oplus \s$ of  spin $j_1$ with respect to the f\/irst copy
 of $\s$ and of spin~$j_2$ with respect to the second copy of $\s$, and thus of dimension
$(2j_1+1)(2j_2 +2)$.

Then all the $4$-dimensional representations of $\s \oplus \s$
are:
\begin{gather}
\label{lista-rep} {\cal D}_{(\frac 32,0)},\ {\cal D}_{(1,0)}\oplus
{\cal D}_{(0,0)},\ {\cal D}_{(\frac 12, \frac12)},\ {\cal
D}_{(\frac12, 0)}\oplus {\cal D}_{(0, \frac 12)},\
 {\cal D}_{(\frac12, 0)}\oplus {\cal D}_{(0,0)}^2,\ {\cal D}_{(0,0)}^4.
\end{gather}
Notice that the third representation in this list corresponds to
the vector representation of the Poincar\'e algebra, the
representation where the momentums $P_m$ also lie. Moreover, it is
also the representation in which the vector generators $V_m$ of
\eqref{algebra} lie. The fourth representation in this list is a
spinor representation, and the supercharges $Q$ of the SUSY
extensions~lie here.

Consider now the $4$ pairs of weights $\alpha_i$, $i=1,\dots,4$
(with respect to  two copies of $\s$) of the $4$ generators of the
representation considered here. The list \eqref{lista-rep} leads
to the following table of weights:
$$
\begin{array}{|c|c|c|c|}
\hline
\vphantom{\Big|}\alpha_1 & \alpha_2 & \alpha_3 & \alpha_4 \\[1mm]
\hline (-3,0) & (-1,0) & (1,0) & (3,0) \\ \hline (-2,0) & (0,0) &
(2,0) & (0,0) \\ \hline
(1,-1) & (-1,1) & (1,1) & (-1,-1) \\ \hline
(1,0) & (-1,0) & (1,0) & (-1,0) \\ \hline
(1,0) & (-1,0) & (0,0) & (0,0) \\ \hline (0,0) & (0,0) & (0,0) &
(0,0) \\ \hline
\end{array} $$

\subsection{Formal connection with the SUSY supercharges}

We make here a formal connection between the bracket for the
generators $V$ (bracket def\/ined in equation \eqref{3-bracket})
and the usual anticommutator of two supercharges $Q$. This
approach is inspired by the fact that, as already noticed in
Section~\ref{sec2}, the SUSY supercharges $Q$ are square roots of
translations and the generators $V$ are cubic roots of
translations. Hence one may think of a~possible connection $
V^3 \propto P \propto Q^2. $ More exactly, recall from the \3
algebra \eqref{algebra} the relation
\begin{gather}
\label{prima} \left\{V_m, V_n, V_r \right \}= \eta_{m n} P_r +
\eta_{m r} P_n + \eta_{r n} P_m.
\end{gather}
The SUSY algebra (with the conventions of \cite{sohnius}) includes
the relation:
\begin{gather}
\label{adoua} \{ Q_\alpha, \bar Q_{\dot \beta} \}= 2
(\sigma^m)_{\alpha \dot \beta} P_m
\end{gather}
(with $\alpha, \dot \beta =1,2$). Hence one may calculate some
connection between the $3$-entries bracket~\eqref{prima} and the
anticommutator~\eqref{adoua}. Writing explicitly the SUSY algebra
anticommutators \eqref{adoua}, one can then express any momentum
$P_m$ as a function of them; these expressions are then inserted
in~\eqref{prima} thus leading to
\begin{gather}
\label{tanasa} \left\{V_m, V_n, V_r \right \} =
(\sigma_{mnr})^{\alpha \dot \beta} \{ Q_\alpha, \bar Q_{\dot
\beta} \},
\end{gather}
where the matrices $\sigma_{mnr}$ have the explicit form
\eqref{matricele}.
\newpage

The purpose of such  connection with the SUSY supercharges would
be to acquire some information on what would be the ``allowed''
possibilities for a vector generator which extends the Poincar\'e
symmetry. However, this does not seem to be the case here, because
in all generality (that is not considering only Lie superalgebras)
the SUSY supercharges $Q$ are not the only fermionic generators to
extend the Poincar\'e symmetries. Indeed, this is the case, as
proved by the Haag--Lopuszanski--Sohnius theorem, but only within
the framework of Lie superalgebras. Considering other type of
algebraic structures, other type of fermionic generators may {\it
a priori} lead to non-trivial extensions and hence relations of
type \eqref{tanasa} could be obtained.

\subsection{Connection with other constructions}
\label{conections}

Already in the introduction we have mentioned other extensions of
the Poincar\'e symmetry. Some other constructions are now referred
to.

In \cite{plus1}, G.~Grignani {\it et.~al.} propose a model for
planar $P,T$-invariant fermions. In this model, a hidden $N=3$
SUSY leads to a non-standard super-extension of the
$(2+1)$-dimensional Poincar\'e group. In \cite{plus2}, K.S.~Nirov
and M.S.~Plyushchay investigate hidden symmetries of a
$P,T$-invariant system of topologically massive $U(1)$ gauge
f\/ields. This system realizes an irreducible representation of a
non-standard super-extension of the $(2+1)$-dimensional Poincar\'e
group. The non-standard behaviour means  that the anticommutator
of corresponding supercharges  results in an operator dif\/ferent
from the Hamiltonian.

In the context of parabosonic systems, in \cite{plus3}, M.S.
Plyushchay founds a hidden polynomial SUSY. One may wonder whether
or not a relation between this structure and the structure
investigated in this review can be established. This seems to be
unlikely, since, even if non-linear, the algebraic structure of
\cite{plus3} is a superalgebra, which is not the case here. Thus,
the two structure are already dif\/ferent at the level of the
def\/inition of the algebraic structure. However, one may further
address the issue of a possible formal connection, like the one
established in the previous subsection for the case with the SUSY
algebra. This still does not seem to be the case for the
non-linear superalgebra appearing in \cite{plus3}, because one
cannot easily obtain  the momentum as function of the
anticommutators, as was done in Appendix~A for the case of the
SUSY algebra.

Another interesting issue is related to the construction of
\cite{plus4}, where  a relativistic wave equation involving the
cubic root of the Klein--Gordon operator is def\/ined. The
equation considered here can be related to fractional
supersymmetry in the following sense: one considers localised
$1$-dimensional fractional supersymmetry, fractional supergravity;
quantising this model one obtains this new type of wave equation
(see~\cite{FSUSY5}).

\subsection{The assumption of analicity}
\label{analicitate}

We have seen in section $1$ that one of the assumptions of the
Coleman--Mandula theorem was the analytic dependence of the
elastic scattering amplitude on the momentum and spin variables.
As S. Coleman and J.~Mandula say in their original paper
\cite{cm}, the naturalness of this assumption being above any
doubt, being ``something that most physicist believe to be a
property of the real world''. In this subsection we will look to
this issue in more detail and give an illustration of why
additional exotic symmetries in the Standard Model frame would
violate this assumption. We do this explicitly on a simple example
of two-body scattering. We then discuss what this becomes for the
case of SUSY and of the algebra \eqref{algebra}. This short
discussion is drawn upon E.~Witten's analyse in \cite{witten}.

In the simplest case of  two-particle scattering, considering that
the momentum and angular momentum are the only conserved charges,
one has the cross section depending only on the scattering angle
$\theta$.

Let us now assume that one has an additional conserved charge, say
a  symmetric, traceless tensor $Z_{mn}$, which closes with the
rest of the Poincar\'e generators by commutation relations. These
commutation relations are not trivial  and hence $Z_{mn}$ has no
trivial matrix elements between particles of dif\/ferent
four-momentum and spin.

For simplicity, we consider here only spinless particle states.
By Lorentz invariance, one takes for the matrix element of
$Z_{mn}$ in a one-particle state of momentum $p$
\begin{gather}
\label{matrix-element} \langle p|Z_{mn}|p\rangle =p_mp_n-\frac 14
\eta_{mn} p^2.
\end{gather}
 (The new matrix element is expressed with the help of the momentum four-vector.)
Now, for the two-particle scattering above, assume that the matrix
element in the two-particle state $|p_1p_2\rangle $ is the sum of
the matrix elements in the states $|p_1\rangle$ and $|p_2\rangle
$. Hence, the conservation of $Z$, $\langle
p_1p_2|Z_{mn}|p_1p_2\rangle =\langle
p'_1p'_2|Z_{mn}|p'_1p'_2\rangle $, leads  to $p_{1\, m}p_{1\,
n}+p_{2\, m}p_{2\, n}=p'_{1\, m}p'_{1\, n}+p'_{2\, m}p'_{2\, n}.$
This is a~supplementary conservation law, which can be satisf\/ied
only for $\theta=0$, and hence the analytic dependence is lost.

If one considers now the case of SUSY, one cannot make the same
reasoning. Indeed, since $Q$ is a spinor, one cannot construct a
matrix element $\langle p|Q_\alpha|p\rangle $ (the analogous of
\eqref{matrix-element}) with the help of only momentum and spin
variables.

To conclude this section, let us now shortly discuss what would
this become for the case of the construction \eqref{algebra}. As
we have already mentioned the additional symmetries $V_m$ are
Lorentz vectors and close within a structure of Lie algebra of
order $3$. Hence, one can {\it a priori} construct a matrix
element like, for example
 $\langle p|V_m|p\rangle = C\, p_m$,
which will lead to a certain constraint. This gives rise to a new,
exotic conservation law which, as above, would contradict analytic
dependence; the theory constructed upon in $4$ dimensions would
have to be non-interacting.

This short remark is an indication towards  conclusion on
impossibility of interactions for the model discussed in this
review. This goes along the result we prove in
Section~\ref{interactiuni}.

\section{Representations of the algebra, bosonic multiplets}\label{sec4}

We have so far introduced the algebra \eqref{algebra} and argued
on some of the foundations of this type of exotic approach. In
this section we start  construction of the f\/ield theoretic model
as follows. Irreducible matrix representations are exhibited.
These lead to dif\/ferent type of multiplets (bosonic or
fermionic) amongst which we consider here the bosonic ones.
Transformation laws of these bosonic f\/ields are then obtained.
We then give some technical properties of these multiplets, the
properties that will prove to be useful for the sequel. Before
ending this section we stop for a moment from the implementation
of our f\/ield theoretical moment and we make some comments
regarding the relation with the spin-statistics theorem.

\subsection{Irreducible representations}

The algebra (\ref{algebra}) has two inequivalent $6$-dimensional
representations
\begin{gather}
\label{matirred} V_{+}{}_m=\begin{pmatrix} 0&\Lambda^{1/3}
\sigma_m& 0 \cr
                           0&0&\Lambda^{1/3} \bar \sigma_m \cr
                            \Lambda^{-2/3}\partial_m&0&0
\end{pmatrix}, \qquad
V_{-}{}_m=\begin{pmatrix} 0& \Lambda^{1/3} \bar \sigma_m& 0 \cr
                           0&0&\Lambda^{1/3}\sigma_m \cr
                            \Lambda^{-2/3} \partial_m&0&0
\end{pmatrix}
\end{gather}
with   $\sigma^m=(\sigma^0=1,\sigma^i)$, and $\bar \sigma^m =(\bar
\sigma^0=1,-\sigma^i)$,  $\sigma^i$ the Pauli matrices and
$\Lambda$ a parameter with mass dimension   that we take  equal to
$1$ (in appropriate units).  For details on the manner this
representation has been obtained, one can refer
to~\cite{articol}. These two representations are referred to as
conjugated to each other and they will give rise to dif\/ferent
types of multiplets, as we will see later on.

One may also notice here that these representations are not proven
to be the only irreducible representations. Thus, if other
representations exist then they may lead also to physics results.

\subsection{Multiplets}
\label{sec-multi}

The matrices $V_+$ and resp. $V_-$ (\ref{matirred}) act on the
triplets of Weyl spinors $\Psi_+$ and resp. $\Psi_-$,
\begin{gather}
\label{spinor} \Psi_+= \begin{pmatrix} \psi_{1 +} \cr  \psi_{2-}
\cr  \psi_{3+} \end{pmatrix}\qquad \mbox{or} \qquad
\Psi_-=\begin{pmatrix}  \psi_{1-}  \cr  \psi_{2+} \cr   \psi_{3-}
\end{pmatrix},
\end{gather}
where $\psi_{1 +}$ is a left-handed (LH) $2$-component Weyl
spinor, $\bar \psi_{2-}$ is a right-handed (RH) $2$-com\-ponent Weyl
spinor {\it etc.}

The transformation law associated to this type of new symmetry
writes
\begin{gather}
\label{3-transf} \delta_v \Psi_\pm=v^m V_{\pm m} \Psi_\pm,
\end{gather}
where $v$ is the transformation parameter.

The representation space of the algebra \eqref{algebra} is
constituted of the states generated by $\Psi$. A very interesting
remark to be made here is that this representation space has a
${\mathbb  Z}_3$-graded structure (the representation space for
the SUSY algebra having a similar ${\mathbb  Z}_2$-graded
structure). Technically, this result comes from the dimension of
the irreducible representation, $6=3\cdot 2$: the $6$-dimensional
matrices $V_m$ act on $3$ copies of Weyl spinors (which are each
of them $2$-dimensional).

In further use, we call the states of $\psi_1$ states of gradation
$-1$, the states of $\psi_2$ states of gradation $0$ and the
states of $\psi_3$ states of gradation $1$.

One has further similitude with SUSY, in the sense already
mentioned of ``cubic roots of translations'': one has
$\psi_1\to\psi_2\to \psi_3\to\psi_1$ (that is, acting with the
generator $V$ on a state of gradation~$-1$ one has a~state of
gradation $0$, acting again with a generator $V$ one has a~state
of gradation~$1$ and f\/inally, acting one more time with a
generator $V$, one has a~state of gradation~$-1$). One could have
reached the same f\/inal state just by acting with some
translation generator on the initial state (recall that a similar
phenomena happened in the case of SUSY, see for example
\cite{sohnius}). Some comments regarding this issue and the
relation to the spin-statistics theorem are made in the last
subsection of this section.

Taking now into consideration the vacuum, a singlet for this new
symmetry, one has to specify in which representation of the
Lorentz algebra it lies. Considering it in the trivial
representation of the Lorentz algebra, equation \eqref{spinor}
leads directly to two fermionic multiplets. We will not be
concerned here with these fermionic multiplets; they have been
treated in \cite{articol} where a non-conventional  kinetic
Lagrangian was obtained.

Another possibility for the vacuum, a Lorentz vector, was also
treated in \cite{articol} leading to the same type of results.

However, considering now  the vacuum lying in the spinor
representation of the Lorentz algebra, one obtains {\it bosonic
multiplets}. The vacuum can be a LH or RH Weyl spinor, $\Omega_+$
and $ \Omega_-$. Therefore one has four possibilities
${\boldsymbol \Xi}_{\pm \pm}$ for a tensor product
$\Psi_\pm\otimes\Omega$, with $\Psi_\pm$ given in \eqref{spinor}
\begin{gather}
{\boldsymbol \Xi}_{++}=\Psi_+ \otimes \Omega_+ = \begin{pmatrix} \Xi_{1++} \\
  \Xi_{2-+} \\ \Xi_{3++} \end{pmatrix}, \qquad
{\boldsymbol \Xi}_{--}=\Psi_- \otimes  \Omega_- =
\begin{pmatrix} \bar \Xi_1{}_{--} \cr
 \Xi_2{}_{+-} \cr \bar
\Xi_3{}_{--}
\end{pmatrix},\nonumber\\
{\boldsymbol \Xi}_{-+}=\Psi_- \otimes  \Omega_+ =
\begin{pmatrix} \Xi_1{}_{-+} \cr
  \Xi_2{}_{++} \cr \Xi_3{}_{-+}
 \end{pmatrix}, \qquad
{\boldsymbol \Xi}_{+-}=\Psi_+ \otimes  \Omega_- =
\begin{pmatrix}  \Xi_1{}_{+-} \cr
 \Xi_2{}_{--} \cr
\Xi_3{}_{+-}
\end{pmatrix}.\label{4-produse}
\end{gather}

The following step is  decomposition of these products of spinors
on $p$-forms. For the case of~${\boldsymbol \Xi}_{++}$, def\/ined
in equation \eqref{4-produse}),
 this writes  (see  \cite{teza} for technical details)
\begin{gather}
\label{muie1832} \Xi_{1++}=\p+\frac14 B_{mn}\sigma^{mn},\qquad
\Xi_{2-+}=\tA_m \bar \sigma^m,\qquad \Xi_{3++}=\ttP+\frac14
\ttB_{mn}\sigma^{mn},
\end{gather}
where we denote by $\p$, $\ttP$ two scalar f\/ields, $\tA_m$ a
vector and $B_{mn}$, $\ttB_{mn}$ two self-dual $2$-forms.

Applying this analysis for the four product of spinors
\eqref{4-produse} leads to the four multiplets ${\boldsymbol
\Xi}_{\pm \pm}$ with the following f\/ield content
\begin{gather}
{\boldsymbol \Xi}_{++}=
 \begin{pmatrix}
\varphi, B_{mn} \cr \tA_m \cr \ttP, \ttB_{mn}
\end{pmatrix}, \qquad
{\boldsymbol \Xi}_{+-}=
\begin{pmatrix}    A'_m \cr \tP', \tB'_{mn} \cr  \ttA'_m
\end{pmatrix},\nonumber\\
{\boldsymbol \Xi}_{--}=
\begin{pmatrix}
\varphi', B^\prime_{mn} \cr \tA'_m \cr \ttP',
\ttB'_{mn}\end{pmatrix},\qquad {\boldsymbol \Xi}_{-+}=
\begin{pmatrix} A_m \cr \tP, \tB_{mn} \cr  \ttA_m \end{pmatrix},\label{4-decomposition}
\end{gather}
where  $\varphi$, $\ttP$, $\varphi'$, $\ttP'$, $\tP$, $\tP'$ are
scalars f\/ields, $\tA$, $\tA'$, $A$, $\ttA$, $A'$, $\ttA'$ are
vector f\/ields, $B$, $\tB$, $\ttB$, $B'$, $\tB'$, $\ttB'$ are
$2$-forms. As we have mentioned above, these $2$-forms, namely
$B$, $\tB$, $\ttB$  are self-dual ({\it i.e.}\ ${}^*B=iB$, where
by ${}^*B$ we mean the dual of $B$) and resp. $B'$, $\tB'$,
$\ttB'$ are anti-self-dual ({\it i.e.}\
 ${}^*B'=-iB'$); thus these $2$-forms must be complex.
To have minimum f\/ield content, one takes ${\boldsymbol
\Xi}_{++}={\boldsymbol \Xi}_{--}^*$ and ${\boldsymbol
\Xi}_{+-}={\boldsymbol \Xi}_{-+}^*$ (that is
$\varphi=\varphi^{'*}$, $B=B^{'*}$, {\it etc.}) We call the
couples ${\boldsymbol \Xi}_{++}-{\boldsymbol \Xi}_{--}$,
${\boldsymbol \Xi}_{+-}- {\boldsymbol \Xi}_{-+}$ {\it conjugated
multiplets}
 and the couples ${\boldsymbol \Xi}_{++}-{\boldsymbol \Xi}_{+-}$,
 ${\boldsymbol \Xi}_{--}- {\boldsymbol \Xi}_{-+}$ {\it interlaced multiplets}.

Following the convention def\/ined earlier, one can say, for
example for the multiplet ${\boldsymbol \Xi}_{++}$ that the
f\/ields $\varphi$, $B$ are of gradation $-1$, the f\/ield $\tA$
is of gradation $0$ and the f\/ields $\ttP$, $\ttB$ are of
gradation $1$.

At this level of our construction let us notice that in the same
multiplet one has scalar and vector f\/ields as well as $2$-form.
In SUSY models, the scalar f\/ields combine with fermions making
supermultiplets. Here, as noted before, since the generators $V$
lie in the vector representation of the Lorentz algebra, one has
multiplets of the same statistics, that is either bosonic or
fermionic multiplets.  Later on in this section we obtain
Lagrangians presenting explicit gauge f\/ixation terms for the
f\/ields. Moreover, in Subsection~\ref{joaja-abeliana},
compatibility of this new type of symmetry with the Abelian gauge
invariance is analysed. This means that, if one applies a gauge
transformation to some multiplet, the same type of multiplet is
obtained (this property being also present for SUSY models).

\subsection[Transformation laws of the fields]{Transformation laws of the f\/ields}
\label{4-transf}

The transformation laws of the f\/ields are obtained from the
transformation law (\ref{3-transf}), using the explicit form
\eqref{matirred} of the matrices $V_{\pm\, m}$ (see again
\cite{teza} for detailed calculus). We give the explicit formulae
obtained for the ${\boldsymbol \Xi}_{++}$ multiplet, similar
formulae being obtained for the other multiplets:
\begin{gather}
\delta_v \varphi = v^m \tA_m,   \qquad \delta_v B_{mn} = - ( v_m
\tA_n - v_n \tA_m )
+ i \varepsilon_{mnpq} v^p \tA^q{}, \nonumber\\
\delta_v \tA_m =  ( v_m \ttP + v^n  \ttB_{mn} ), \qquad \delta_v
\ttP = v^m
\partial_m \varphi, \qquad
\delta_v  \ttB_{mn} = v^p \partial_p B_{mn}. \label{transfo2}
\end{gather}

\subsection{Derivation of a multiplet}
\label{deriv-mult}

We now obtain an interesting property of these multiplets,
property which will be used when analysing  compatibility of our
model with Abelian gauge invariance (see Subsection
\ref{joaja-abeliana}) and when treating the possibilities of
interaction (see Section~\ref{interactiuni}).

Let us denote from now on by  $X_{[mn]\pm}$  the
(anti-)self-dualisation of any second rank tensor~$X_{mn}$, {\it
i.e.}\ $
X_{[m n]_\pm} = X_{m n} - X_{n m} \mp i \varepsilon_{mnpq} X^{p
q}. $

Let us now consider the f\/ields of a ${\boldsymbol \Xi}_{+-}$
multiplet, that is $ A'_m$; $\tP'$, $\tB'_{mn}$; $\ttA'_m$ to
construct a~dif\/ferent type of multip\-let using partial
derivatives~$\partial_m$. Thus, to construct a ${\boldsymbol
\Xi}_{++}$ multip\-let, one has to have expressions for any f\/ield
of the ${\boldsymbol \Xi}_{++}$ multiplet. Saturating the Lorentz
indices, respecting the $\ZZ_3$ gradation and the (anti-)self-dual
character of the dif\/ferent $2$-forms,
 one possible solution is
\begin{gather}
{\cal D} \,{\boldsymbol \Xi}_{+-} =
\Big(\psi, \psi_{mn}, \; \tpsi_m, \; \ttpsi, \ttpsi_{mn} \Big) \nonumber \\
\phantom{{\cal D} \,{\boldsymbol \Xi}_{+-}}{}
 \equiv  \Big(\partial_m A'^m, \partial_{[m}A'_{n]_+}; \; \partial_m \tP'+
\partial^n \tB'^{}_{nm}; \; \partial_m \ttA'^m, \partial_{[m}\ttA'_{n]_+}\Big). \label{partialXipm}
\end{gather}
The last thing  for this set to form a ${\boldsymbol \Xi}_{++}$
multiplet is that it transforms  as requested by
equation~\eqref{transfo2}. This is checked by directly applying
the transformations laws~\eqref{transfo2} on
equation~\eqref{partialXipm}. Thus we have shown a mechanism to
obtain a multiplet of a certain type (here ${\boldsymbol
\Xi}_{++}$) by ``deriving'' a multiplet of another type (here
${\boldsymbol \Xi}_{+-}$). We call ${\cal D} {\boldsymbol
\Xi}_{+-}$ a {\it derivative multiplet}. One can actually def\/ine
such a ``derivation'' for every  \3 multiplet ${\boldsymbol
\Xi}_{\pm\pm}$.

\subsection{The spin-statistics connection}

Before going further in developing this model, we would like to
address here the legitimate question of the connection with the
spin-statistics theorem, theorem which states that bosons obey the
Bose--Einstein statistics and fermions obey the Fermi--Dirac
statistics. So one may ask where does our model stay from this
point of view?

The additional symmetries $V$ even though they lie in the vector
representation of the Lorentz algebra do not close with classical
(anti)commutation relations. Usually in physics literature (for
example in the case of superalgebras), one denotes generators that
close with commutators as bosonic generators and to generators
that close with anticommutators as fermionic generators. Obviously
this is not the case here: $V$ are neither bosonic nor fermionic
generators (this is how our construction evades the no-go theorems
and apparently, one of the prices to pay). Nevertheless, the
physical f\/ields $\varphi$, $A$, $B$ are bosons (and in the case
of the fermionic multiplets in \cite{articol} they are fermions),
thus obeying the conventional statistics. Technically, we got to
this situation by decomposition on $p$-forms.

A connected aspect here is the following. In the SUSY case, one
has a $\ZZ_2$-graded structure which, at the level of the
representation space translates by a division of the
representation space in two subspaces, a bosonic and a fermionic
subspace. This interpretation is obviously lost here: we have  now
a $\ZZ_3$-graded representation space and we do not have a
correspondence of these subspaces to some type of particle, as was
the case for SUSY.   However, as mentioned above, one f\/inds the
conventional types of f\/ields at the level of the physical
f\/ields $\varphi$, $A$, $B$ {\it etc.}

\section{Free theory}
\label{4-liber}

In this section we construct free Lagrangians invariant under the
transformations \eqref{transfo2}; one thus obtains a {\it new
symmetry}. From now on we denote the f\/ield strengths associated
to the f\/ields by $F_{mn}=\partial_m A_n-\partial_n A_m$ for any
vector f\/ield $A_m$ and by $H_{mnp}=\partial_m B_{np}+\partial_p
B_{mn}+\partial_n B_{pm}$ for any $2$-form $B_{mn}$.

\subsection{Coupling between conjugated multiplets}
\label{detoate}

If we consider the quadratic couplings between conjugated
multiplets,
 as denoted in Subsection~\ref{sec-multi}, one can construct two invariant  Lagrangians, one for each pair
${\boldsymbol \Xi}_{++}-{\boldsymbol \Xi}_{--}$ and ${\boldsymbol
\Xi}_{+-}-{\boldsymbol \Xi}_{-+}$
\begin{gather}
{\cal L}_0={\cal L}_0({\boldsymbol \Xi}_{++}) +{\cal L}_0({\boldsymbol \Xi}_{--} ) \nonumber \\
\phantom{{\cal L}_0}{} =\partial_m \varphi \partial^m \ttP +
\frac{1}{12} H_{mnp} {\tilde {\tilde H}}^{mnp} +\frac12 {}^\star
H_m {}^\star {\tilde {\tilde H}}^m -\frac14 \tilde F_{mn} \tilde
F^{mn}
-\frac12 \big(\partial_m \tilde A^m\big)^2 \nonumber \\
\phantom{{\cal L}_0=}{}+ \partial_m \varphi' \partial^m \ttP' +
\frac{1}{12} H'_{mnp} {\tilde {\tilde H}}^{\prime mnp} +\frac12
{}^\star H'_m {}^\star {\tilde {\tilde H}}^{\prime m} -\frac14
\tilde F'_{mn} \tilde F'^{mn} -\frac12 \big(\partial_m \tilde
A'^m\big)^2\label{free4}
\end{gather}
and
\begin{gather}
{\cal L'}_0={\cal L}_0({\boldsymbol \Xi}_{-+}) +{\cal L}_0({\boldsymbol \Xi}_{+-} ) \nonumber \\
\phantom{{\cal L'}_0}{}=\frac12 \partial_m \tP \partial^m \tP
+\frac{1}{24} \tilde H_{mnp} \tilde H^{mnp} - \frac14  {}^\star
\tilde H_m {}^\star \tilde H^m -\frac12 F_{mn} {\tilde {\tilde
F}}^{mn} -
(\partial_m A^m)\big(\partial_n \ttA^n\big)  \nonumber \\
\phantom{{\cal L'}_0=}{}+\frac12 \partial_m \tP' \partial^m \tP'
+\frac{1}{24} \tilde H'_{mnp} \tilde H'^{mnp} - \frac14  {}^\star
\tilde H'_m {}^\star \tilde H'^m -\frac12 F'_{mn} {\tilde {\tilde
F'}}^{mn} - (\partial_m A'^m)\big(\partial_n
\ttA'^n\big).\!\!\!\label{4free2}
\end{gather}
We consider here ${\cal L}_0$, the analyse for ${\cal L}'_0$ being
identical. Here we have denoted ${}^*H_m=\frac16 \e_{mnpq}
H^{npq}=\partial^n B_{mn}$ the dual of the f\/ield strength $H$.
Since we use complex conjugated terms, the Lagrangian
\eqref{free4} is real; furthermore, it is of gradation $0$.

Let us now perform  in \eqref{free4} the following change of
variables
\begin{gather}
\tA_1=\frac{\tA+\tA^\prime}{\sqrt{2}},\qquad \tA_2=
i\frac{\tA-\tA^\prime}{\sqrt{2}}, \qquad B_1=
\frac{B+B^\prime}{\sqrt{2}},\qquad
B_2= i\frac{B-B^\prime}{\sqrt{2}}, \nonumber \\
\ttB_1=  \frac{\ttB+\ttB^\prime}{\sqrt{2}} ,\qquad \ttB_2=
i\frac{\ttB- \ttB^\prime}{\sqrt{2}}, \qquad \varphi_1=
\frac{\varphi+\varphi^\prime}{\sqrt{2}} , \qquad
\varphi_2= i\frac{\varphi-\varphi^\prime}{\sqrt{2}}, \nonumber \\
\ttP_1=  \frac{\ttP+\ttP^\prime}{\sqrt{2}} , \qquad \ttP_2=
i\frac{\ttP- \ttP^\prime}{\sqrt{2}}.\label{real}
\end{gather}
The Lagrangian writes now
\begin{gather*}
{\cal L}_0 =\partial_m \varphi_1 \partial^m \ttP_1 -
 \partial_m \varphi_2 \partial^m \ttP_2\nonumber\\
\phantom{{\cal L}_0=}{}+ \frac{1}{6} H_1{}_{mnp} {\tilde {\tilde
H}}_1^{mnp} + \partial^n B_1{}_{nm} \partial_p \ttB_1{}^{pm} -
\frac{1}{6} H_2{}_{mnp} {\tilde {\tilde H}}_2^{mnp} - \partial^n
B_1{}_{nm} \partial_p \ttB_1{}^{pm}
\nonumber \\
\phantom{{\cal L}_0=}{}-\frac14 F_1{}_{mn}  F_1{}^{mn} +\frac14
{\tilde F}_2{}_{mn} {\tilde F}_2{}^{mn} -\frac12 \big(\partial_m
\tilde A_1{}^m\big)^2
+\frac12 \big(\partial_m \tilde A_2{}^m\big)^2. 
\end{gather*}
Notice at this point that by the redef\/inition \eqref{real}, we
f\/ind ourselves with $2$-forms $B_1$, $B_2$, $\ttB_1$
and~$\ttB_2$ which are neither self-dual nor anti-self-dual.
Moreover, one observes
 ${}^\star B_1 = B_2$, ${}^\star \ttB_1 = \ttB_2$.
Therefore,
 one can eliminate two of them, for example $B_2$ and $\ttB_2$; thus ${\cal L}_0$  now becomes
\begin{gather}
{\cal L}_0 =\partial_m \varphi_1 \partial^m \ttP_1 -
 \partial_m \varphi_2 \partial^m \ttP_2
+ \frac{1}{6} H_1{}_{mnp} {\tilde {\tilde H}}_1^{mnp} + \partial^n
B_1{}_{nm} \partial_p \ttB_1{}^{pm}
\nonumber \\
\phantom{{\cal L}_0=}{} -\frac14 F_1{}_{mn}  F_1{}^{mn} +\frac14
{\tilde F}_2{}_{mn} {\tilde F}_2{}^{mn} -\frac12 \big(\partial_m
\tilde A_1{}^m\big)^2 +\frac12 \big(\partial_m \tilde
A_2{}^m\big)^2.\label{free4_2}
\end{gather}

Proceeding with the analysis of this Lagrangian, one notices that
the terms in the f\/irst line of (\ref{free4_2}) are not diagonal.
For this purpose, we now def\/ine
\begin{gather}
\hat \varphi_1 = \frac{\varphi_1 + \ttP_1}{\sqrt{2}},\qquad {\hat{
\hat \varphi}}_1 = \frac{\varphi_1 - \ttP_1}{\sqrt{2}},\qquad \hat
\varphi_2 = \frac{\varphi_2 + \ttP_2}{\sqrt{2}},\qquad
{\hat{ \hat \varphi}}_2 = \frac{\varphi_2 - \ttP_2}{\sqrt{2}}, \nonumber \\
\hat B_1 = \frac{B_1 + \ttB_1}{\sqrt{2}},\qquad {\hat{ \hat B}}_1
= \frac{B_1 - \ttB_1}{\sqrt{2}},\label{diag}
\end{gather}
and, with the new f\/ields,  ${\cal L}_0$ writes
\begin{gather}
 {\cal L}_0 =\frac12 \partial_m \hat \varphi_1 \partial^m  \hat \varphi_1
- \frac12 \partial_m {\hat {\hat \varphi}}_1
\partial^m  {\hat {\hat \varphi}}_1 -
\frac12 \partial_m \hat \varphi_2 \partial^m  \hat \varphi_2 +
\frac12 \partial_m {\hat {\hat \varphi}}_2
\partial^m  {\hat {\hat \varphi}}_2 \nonumber \\
\phantom{{\cal L}_0 =}{}-\frac14 \tilde F_1{}_{mn} \tilde
F_1{}^{mn} +\frac14 \tilde F_2{}_{mn} \tilde F_2{}^{mn} -\frac12
\big(\partial_m \tilde A_1{}^m\big)^2
+\frac12 \big(\partial_m \tilde A_2{}^m\big)^2 \nonumber \\
\phantom{{\cal L}_0 =}{}+\frac16 \hat H_1{}_{mnp} \hat H_1^{mnp} +
\partial^n \hat B_1{}_{nm}\partial_p \hat B_1^{pm} -\frac16 {\hat
{\hat H}}_1{}_{mnp} {\hat {\hat H}}_1^{mnp} - \partial^n
\hhB_1{}_{nm}\partial_p \hhB_1^{pm}.\label{fermi2}
\end{gather}

The above Lagrangian writes also as
\begin{gather}
 {\cal L}_0 =
\frac12 \partial_m \hat \varphi_1 \partial^m  \hat \varphi_1 -
\frac12 \partial_m {\hat {\hat \varphi}}_1
\partial^m  {\hat {\hat \varphi}}_1 -
\frac12 \partial_m \hat \varphi_2 \partial^m  \hat \varphi_2 +
\frac12 \partial_m {\hat {\hat \varphi}}_2
\partial^m  {\hat {\hat \varphi}}_2 \nonumber \\
\phantom{{\cal L}_0 =}{}-\frac12 \partial_m \tA_1{}_n \partial^m
\tA_1{}^n +\frac12 \partial_m \tA_2{}_n \partial^m \tA_2{}^n
+\frac14 \partial_m \hat B_1{}_{np} \partial^m \hat B_1{}^{np}
 -\frac14 \partial_m {\hat {\hat B}}_1{}_{np} \partial^m {\hat {\hat B}}_1{}^{np}. \label{fermi}
\end{gather}
This form is actually suitable for further diagonalisation
computations (see Subsection~\ref{diagonalizarea-mare}).

Let us now consider the general gauge transformations
\begin{gather}
\label{joaja-normala} A_m\to A_m + \partial_m \chi,\qquad
B_{mn}\to B_{mn}+\partial_m \chi_n-\partial_n \chi_m,
\end{gather}
where $\chi$ and $\chi_m$ are the gauge parameters. Hence one sees
in the Lagrangian  \eqref{fermi2}  presence of kinetic terms and
Feynman gauge f\/ixing terms (of type $-\frac12 \left(\partial_m
A^m\right)^2$ for a generic vector f\/ield~$A$ or of type
$\partial^n B_{nm}\partial_p B^{pm}$ for a generic $2$-form $B$).
These gauge f\/ixing terms are not just a {\it choice} of gauge,
but they are {\it required by the invariance}.

The gauge f\/ixing terms above (of type $-\frac12 \left(\partial_m
A^m\right)^2$ and resp.\ $\partial^n B_{nm}\partial_p B^{pm}$)
imply some constraints on the gauge parameters def\/ined in
\eqref{joaja-normala}, namely
\begin{gather}
\label{contrainte-joaja}
\partial^m \partial_m \chi= 0,\qquad
\partial^m (\partial_m \chi_n-\partial_n \chi_m)=0.
\end{gather} We will come back on this issue of gauge transformation in Subsection~\ref{joaja-abeliana}.
The presence of these gauge f\/ixing terms has a lot of
consequences on dif\/ferent aspects of the model.

The f\/irst of them is related to the number of degrees of freedom
of our f\/ields. A $p$-form in $D$ dimensions has
 $C_D^p$
independent components. Let us now introduce for simplif\/ication
dif\/ferential forms notations, namely $A_{[p]}$ for a generic
$p$-form, $d$ for the exterior derivative (mapping a $p$-form into
a $(p+1)$-form) and for  $d^\dag$
 for its adjoint (mapping a $p$-form into a $(p+1)$-form).

If one deals with a generic {\it free} $p$-form $\omega_{[p]}$,
then the gauge transformation is
\begin{gather*}
 \omega_{[p]}\to \omega_{[p]} + d \chi_{[p-1]}, \nonumber
\end{gather*}
where the gauge parameter $\chi_{[p-1]}$ is a $(p-1)$-form. Using
the reducibility character of the gauge transformation, one
obtains at the end degrees of freedom for such a free  of\/f-shell
$p$-form
 (see for example \cite{gomis}).

Consider for example the well-known case of a photon, a $1$-form,
in four dimensions, we have $C_3^1=3$ degrees of freedom. To
f\/ind the well-known number of $2$ degrees of freedom for a {\it
physical} photon, one uses Ward identities (for a detailed
analysis see for example~\cite{peskin}).

For the general case of a on-shell $p$-form ($p\le D-2$) in $D$
dimensions, similar Ward identities lead to
 $ C_{D-2}^p$
 physical degrees of freedom.

This is not the case for the f\/ields here;  the gauge parameters
are subject to constraints of type \eqref{contrainte-joaja};
 thus one cannot anymore eliminate degrees of freedom of the $p$-form as was the case before.
Hence, the $p$-form has $C_D^p$
 degrees of freedom ($4$ for a vector f\/ield and $6$ for a $2$-form).

Another important aspect of the Lagrangian (\ref{fermi}) is that
the f\/ields
\begin{gather}
\label{bolnavele} {\hat {\hat \varphi}}_1,\quad {\hat {
\varphi}}_2, \quad \tA_2, \quad {\hat {\hat B}}_1
\end{gather}
have wrong sign for their kinetic term. This implies {\it a
priori} a problem of unboundedness from below of their potentials.
This problem might by corrected by suitable self-interaction
terms, but as we will see in the following section this is not the
case because this type of terms is not allowed by this symmetry.

A possible solution to the problem of unboundness from below is
based on Hodge duality of $p$-forms. Recall that for the situation
of free forms, dualisation is performed at the level of the
f\/ield strength; this implies  equivalence of the theories of a
$p$-form and a $(D-2-p)$-form (in $D$ dimensions). Indeed,
starting from a generic $p$-form $\omega_{[p]}$, one considers its
f\/ield strength, $d\omega_{[p]}$ which is a $(p+1)$-form.
Considering its Hodge dual, we have a $(D-p-1)$-form, which is the
f\/ield strength of $(D-p-2)$-form. One can see that these
theories are equivalent, having the same number of physical
degrees of freedom $ C_{D-2}^p=C_{D-2}^{D-p-2}$. This can be
written schematically as
\begin{gather*}
\xymatrix{   F_{[p+1]}=d\omega_{[p]} & \cong & ({}^*F)_{[D-p-1]}\ar[d]\\
   \omega_{[p]} \ar[u] & & \omega'_{[D-p-2]}}
\end{gather*}

In \cite{io1}, a dif\/ferent type of dualisation is proposed for
our model. This dualisation is performed at the level of the
potential directly and not at the level of its f\/ield strength,
as above. This means that one replaces the f\/ields
\eqref{bolnavele} which have wrong signs in the Lagrangian, by
their Hodge dual f\/ields. Using appropriate identities relating
$p$-forms kinetic terms and their Hodge duals, one has correct
signs for these Hodge dual f\/ields (see \cite{io1} for details).

Finally, notice that because of the gauge f\/ixation present here,
any $p$-form and its Hodge dual (with whom we have replaced in the
Lagrangian the $p$-forms \eqref{bolnavele}) will have the same
number of degrees of freedom $ C_{D}^{D-p}=C_D^p$.

\subsection{Coupling between interlaced multiplets}
\label{embedded}

So far we have analysed invariant terms that arise from couplings
of conjugated multiplets, ${\boldsymbol \Xi}_{++}-{\boldsymbol
\Xi}_{--}$ and ${\boldsymbol \Xi}_{+-}-{\boldsymbol \Xi}_{-+}$. We
now look closer to couplings between the pairs of interlaced
multiplets, ${\boldsymbol \Xi}_{++}-{\boldsymbol \Xi}_{+-}$ and
${\boldsymbol \Xi}_{--}-{\boldsymbol \Xi}_{-+}$. We prove that
quadratic coupling terms between these pairs are also invariant.

 Starting the calculations with the f\/ields given in (\ref{4-decomposition}),
 one can write the following $0$-graded, real Lagrangian
\begin{gather}
{\cal L}_{\mathrm{c}} = {\cal L}_{\mathrm{c}} ({\boldsymbol
\Xi}_{++},{\boldsymbol \Xi}_{+-}) + {\cal
L}_{\mathrm{c}}({\boldsymbol \Xi}_{--},{\boldsymbol \Xi}_{-+})
 \nonumber\\
\phantom{{\cal L}_{\mathrm{c}}}{}=\lambda \big(
\partial_m \varphi \ttA^\prime{}^m
+\partial_m \ttP  A^\prime{}^m -\partial_m \tA^m \tP^\prime -
\partial_m \tA_n \tB^\prime{}^{mn} +\partial^m  B_{mn}
\ttA^\prime{}^n
 +\partial^m \ttB_{mn} A^\prime{}^n
\big)  \nonumber \\
\phantom{{\cal L}_{\mathrm{c}} =}{}+\lambda^\star \big(\partial_m
\varphi' \ttA{}^m +\partial_m \ttP'  A^m -\partial_m \tA'^m \tP -
\partial_m \tA'_n \tB^{mn} +\partial^m  B'_{mn}  \ttA^n
 +\partial^m \ttB'_{mn} A^n\big), \label{coup}
\end{gather}
with $\lambda=\lambda_1 + i \lambda_2$ a complex  coupling
constant with mass dimension.

To study the invariance of (\ref{coup}) one may study separately
the invariance of ${\cal L}_{\mathrm{c}}({\boldsymbol
\Xi}_{++},{\boldsymbol \Xi}_{+-})$ and ${\cal
L}_{\mathrm{c}}({\boldsymbol \Xi}_{--},{\boldsymbol \Xi}_{-+})$
because they do not mix under \3 transformations (\ref{transfo2}).
Up to total derivative, one has $\delta_v{\cal
L}_{\mathrm{c}}({\boldsymbol \Xi}_{++},{\boldsymbol \Xi}_{+-}) = 0
$.

 A few remarks are in order to be made now. First,
 one can check that the gauge f\/ixation of~${\cal L}_0$ and ${\cal L}'_0$
 is still demanded by the terms of ${\cal L}_c$ (the last two lines of equation \eqref{L1/3}).
 Indeed, if one looks at the $2$-form $B$, then terms of type $\frac12 B^{mn} F{}_{mn}$
 f\/ix the gauge, while terms of type ${}^\star B^{mn} F{}_{mn}$
are  gauge invariant. These terms, known as $BF$-terms are related
to topological theories \cite{BF, BF2,BF3,BF4}. Nevertheless, this
line of work is not the one used here (for example we have never
been concerned with surface terms in any of our invariance
calculations).

A last thing to notice is that couplings like $A^m \partial_m
\varphi$ present in \eqref{L1/3} are of Goldstone type. Usually,
they are gauged away and are responsible for  appearance of mass
(see for example~\cite{peskin}). However, one sees that this
mechanism cannot be applied in the case of our model, since, as we
have already stated above, the gauge is partially f\/ixed.

\subsection{Diagonalisation of the total Lagrangian}
\label{diagonalizarea-mare}

The total Lagrangian to be considered is
\begin{gather}
\label{free} {\cal L}_t= {\cal L}_0 + {\cal L}^\prime_0 + {\cal
L}_c,
\end{gather}
where ${\cal L}_0$ and ${\cal L}'_0$ are given in (\ref{free4})
and resp.~\eqref{4free2} and ${\cal L}_c$ is given in
(\ref{coup}). Since ${\cal L}$ is quadratic in the f\/ields, we
deal with a non-interacting theory and it should be possible, by
f\/ield redef\/initions to write the Lagrangian in a diagonal
form.

In order to do this, we f\/irst perform the changes of variable
(\ref{real}) and (\ref{diag}) that make  ${\cal L}_0$ exp\-licitly
real and diagonal. Obviously, the same redef\/initions (keeping
the same type of notations for the redef\/ined f\/ields) must be
made for ${\cal L}'_0$. After all this, the f\/ield content is:
\begin{enumerate}\itemsep=0pt
\item[]$6$ scalar f\/ields,   $\hP_1$, $\hhP_1$, $\hP_2$, $\hhP_2$
(in ${\cal L}_0$), $\tP_1$, $\ttP_2$ (in ${\cal L}'_0$); \item []
$6$ vector f\/ields,   $\tA_1$, $\ttA_2$ (in ${\cal L}_0$),
$\hA_1$, $\hhA_1$, $\hA_2$, $\hhA_2$ (in ${\cal L}'_0$); \item[]
$3$ two-forms $\hB_1$, $\hhB_1$ (in ${\cal L}_0$) $\tB_2$ (in
${\cal L}'_0$).
\end{enumerate}

We thus have a total of $15$ independent f\/ields. Expressed with
these new f\/ields, ${\cal L}_t$ decouples into $3$ distinct
pieces, each of them having the exact same dependence on a set of
$5$ f\/ields (two scalars, two vectors and one $2$-forms, denoted
generically by $\varphi_1$, $\varphi_2$, $A_1$, $A_2$ and $B$).
This Lagrangian writes
\begin{gather}
{\cal L}(\varphi_1, \varphi_2, A_1, A_2, B)= \frac12 (\partial_m
\varphi_1)^2- \frac12 (\partial_m \varphi_2)^2- \frac12
(\partial_m A_1{}_n)^2 +\frac12 (\partial_m A_2{}_n)^2
+\frac14 (\partial_m B_{np})^2 \nonumber \\
\phantom{{\cal L}(\varphi_1, \varphi_2, A_1, A_2, B)=}{}
+\lambda_1\big(A_1{}^m \partial_m \varphi_1 + A_2{}^m \partial_m
\varphi_2
 - B^{mn} \partial_m A_1{}_n - {}^\star B^{mn} \partial_m A_2{}_n
\big)\nonumber\\
\phantom{{\cal L}(\varphi_1, \varphi_2, A_1, A_2, B)=}{}
+\lambda_2\big(-A_2{}^m \partial_m \varphi_1 + A_1{}^m \partial_m
\varphi_2
 + B^{mn} \partial_m A_2{}_n - {}^\star B^{mn} \partial_m A_1{}_n
\big).\!\!\! \label{L1/3}
\end{gather}

Thus, for diagonalising the total Lagrangian ${\cal L}_t$ it is
enough to work on ${\cal L}(\varphi_1, \varphi_2, A_1, A_2, B)$.
Expressing the Lagrangian in the Fourier space, the f\/irst step
is to complete a perfect square for the terms involving $\tA_1$
(we denote in the rest of this subsection the Fourier transforms
by tilde, not to be confused with the tilde in the f\/ields we had
until  equation (\ref{L1/3})). One has to def\/ine
\begin{gather*}
\tA'_1{}_m(p)= \tA_1{}_m(p)  +
 \frac{ \lambda_1}{p^2}i p_m \tP_1(p) + \frac{  \lambda_2}{p^2}i p_m \tP_2(p)
+  \frac{ \lambda_1}{p^2}i p^r\tB_{rm}(p) + \frac{
\lambda_2}{p^2}ip^r ({}^\star \tB_{rm}(p)).
\end{gather*}
The next step is to complete a perfect square for the terms
involving $\tA_2$ (the Fourier transform of the vector f\/ield
$A_1$); one def\/ines
\begin{gather*}
\tA'_2{}_m(p)= \tA_2{}_m(p)  - \frac{ \lambda_1}{p^2}ip_m \tP_2(p)
+ \frac{ \lambda_2}{p^2}ip_m \tP_1(p) +  \frac{
\lambda_2}{p^2}ip^r \tB_{rm}(p) - \frac{ \lambda_1}{p^2}ip^r
 ({}^\star \tB_{rm}(p)).
\end{gather*}

A f\/inal diagonalisation can be written for $\tP$. Thus one
def\/ines
\begin{gather*}
\tP'(p)=\tP(p)+\frac{\lambda_1 \lambda_2}{\frac12
(p^2-(\lambda_2^2-\lambda_1^2))} \tP_2(p).
\end{gather*}
After all this, the Lagrangian f\/inally writes as
\begin{gather}
\tilde {\cal L} =\frac12\left(p^2-(\lambda_2^2-\lambda_1^2)\right)
 \tP'_1(p)\tP'_1(-p)\nonumber\\
 \phantom{\tilde {\cal L} =}{}
               - \frac12\left(p^2-(\lambda_2^2-\lambda_1^2)
               +\frac{\lambda_1^2 \lambda_2^2}{\frac12 (p^2-(\lambda_2^2-\lambda_1^2))}
\right) \tP_2(p)\tP_2(-p)   \nonumber \\
 \phantom{\tilde {\cal L} =}{}- \frac12 p^2 \tA'_1{}_m(p)  \tA'_1{}^m(-p)    + \frac12 p^2 \tA'_2{}_m(p)
 \tA'_2{}^m(-p) +
\frac14    p^2 \tB_{mn}(p)  \tB^{mn}(-p)   \nonumber\\
 \phantom{\tilde {\cal L} =}{}+\frac12 \frac{1}{p^2} p^r p_s(\lambda_1^2 -\lambda_2^2)\left(
\tB_{rm}(p) \tB^{sm}(-p)- {}^\star \tB_{rm}(p) {}^\star
\tB^{sm}(-p)\right)
 \nonumber \\
 \phantom{\tilde {\cal L} =}{}+\frac{\lambda_1 \lambda_2}{p^2}  p^r p_s
\left(\tB_{rm}(p) {}^\star \tB^{sm}(-p)+ {} \tB^{s m}(p) {}^\star
\tB_{rm}(-p)\right).\label{L-diag2}
\end{gather}
One can now remark that not all values of the  parameters
$\lambda_1$ and $\lambda_2$ are allowed if we do not want
tachyons to be present. Some allowed values (which considerably
simplify  the Lagrangian~(\ref{L-diag2})) are
$\lambda_1=\lambda_2$ or $\lambda_1=0$. However one remarks a
non-conventional form of the kinetic term for the $2$-form $B$.
One f\/inal remark is that we have done this diagonalisation on
the Lagrangian without  dualisation mentioned at the end of
Subsection~\ref{detoate}; the sign of the kinetic terms will not
change; however the same type of calculation may be performed for
the dualised Lagrangian.

One more remark is to be made. Even if we have so far considered
non-massive f\/ields, invariant mass terms can be explicitly added
to our Lagrangian (see \cite{io1}). Moreover, as one can easily
see from the \3 algebra \eqref{algebra}, $P^2$ is a Casimir
operator and therefore all states in an irreducible representation
must have the same mass.

The  analysis done so far for the massless multiplets do not
drastically change. For example, in the massive case one has no
gauge invariance and the number of degrees of freedom remains the
same (one does not have gauge parameters to eliminate any degree
of freedom). For  diagonalisation of the total free Lagrangian,
these mass terms do not change  our analysis qualitatively.

\subsection{Compatibility with  Abelian gauge invariance}
\label{joaja-abeliana}

We now look closer to the problem of the compatibility with
Abelian gauge symmetry. We have seen that these two symmetries are
closely connected, in the sense that the gauge symmetry is f\/ixed
by the Feynman gauge f\/ixing terms required in the Lagrangian
(see Subsections~\ref{detoate} and~\ref{diagonalizarea-mare}). So
far we have also seen some of the consequences of this gauge
f\/ixing, such as  possibility of dualisation related to the equal
number of degrees of freedom (see Subsection~\ref{detoate}) and
also  impossibility of gauging away the coupling terms of ${\cal
L}_c$~\eqref{coup}.

Nevertheless, another question is appropriate at this level. If
one acts with the gauge transformation on a multiplet will the
result be a multiplet? Or, schematically, ${\boldsymbol \Xi}
\overset{{\rm gauge}}\longrightarrow {\boldsymbol \Xi'}$? So what
one has to check is whether or not ${\boldsymbol \Xi'}$ is a \3
multiplet. (Recall that this is the case for SUSY, where a gauge
transformation transforms a vector superf\/ield $V$ to
$V+\Phi+\Phi^\dag$, with~$\Phi$ a~chiral superf\/ield (see for
example \cite{sohnius}).) Moreover we also f\/ind in what
conditions  gauge parameters may form a \3 multiplet.

{\it \underline{I}}. Let us f\/irst write the general gauge
transformation one uses for the physical f\/ields $\hat \p_1$,
$\hat \p_2$, $\hhP_1$, $\hhP_2$, $\tA_1$, $\tA_2$, $\hat B_1$,
$\hhB_1$
\begin{gather}
\label{jauge-normale} \hP_1 \to \hP_1 + \hat k_1,\qquad
\tA_{1m}\to \tA_{1m}+\partial_m \tilde \chi_1, \qquad \hB_{1mn}
\to {\hB}_{1mn}+\partial_m \hat \chi_{1n} - \partial_n \hat
\chi_{1m}.
\end{gather}
and similarly for the rest of the f\/ields ($\hat k_1$ being some
constant). Recall that these physical f\/ields were obtained from
linear redef\/initions of the original f\/ields
(\ref{4-decomposition}). Hence one can write down gauge
transformations of the  f\/ields (\ref{4-decomposition}) also. For
example, the vector f\/ield $\tA_1$ was obtained from the f\/ields
$\tA$ and $\tA'$ by the redef\/inition \eqref{real},
$\tA_1=\frac{1}{\sqrt 2}(\tA + \tA')$. Hence the gauge 
parameter~$\tilde \chi_1$ of \eqref{jauge-normale} is written as $\tilde
\chi_1=\frac{1}{\sqrt 2}(\tilde \chi + \tilde \chi')$ that will
thus allow us to obtain the gauge parameters~$\tilde \chi$
and~$\tilde \chi'$ of the f\/ield $\tA$ and respectively~$\tA'$.

The gauge transformations for the original f\/ields
(\ref{4-decomposition}) thus write
\begin{gather} \label{g-trans}
\phi \to \phi + k,  \qquad \tA_m \to \tA_m + \partial_m \tilde
\chi,   \qquad { B}^{(\pm)}_{m n} \to { B}^{(\pm)}_{m n} +
\partial_m \chi_n - \partial_n \chi_m  \mp i \varepsilon_{mnpq} \partial^p \chi^q
\end{gather}
and similarly for the rest of the f\/ields. Note, however, that
for the case of $2$-forms,
 compatibility between the  transformations \eqref{jauge-normale}
and \eqref{g-trans} needs a closer look. Indeed, write in $p$-form
notation the transformation \eqref{g-trans} for the self-dual
$2$-form $B_{mn}$
\begin{gather}
\label{g-trans-forme} {B}^{} \to {B}^{} + d \chi_{[1]}- i {}^*
(d\chi_{[1]}).
\end{gather}
Recall now that $B'=B^*$; one has
\begin{gather}
\label{ttb'} {B'} \to {B'} + d \chi^*_{[1]}+ i {}^*
(d\chi^*_{[1]}).
\end{gather}
(One should pay attention to the notations used for  dual and
complex conjugation, that is ${}^*B$ denotes the dual of $B$
whereas $B^{*}$ denotes the complex conjugated of $B$.) Equation
\eqref{real} combined these two $2$-forms in the real $2$-form
$B_1=\frac{1}{\sqrt 2}(B+B')$. Thus, its gauge transformation
writes
\begin{gather}
\label{b1} B_1\to B_1+ \frac{1}{\sqrt 2}
\big(d(\chi_{[1]}+\chi^*_{[1]})-i{}^* d(\chi_{[1]}
-\chi^*_{[1]})\big).
\end{gather}
One immediate solution for \eqref{b1} to be the gauge
transformation \eqref{jauge-normale} for a $2$-form, is to impose
that the $1$-form $\chi_{[1]}$ is real.

However, we now prove that this compatibility can still be
achieved even if  the $1$-form $\chi_{[1]}$~is complex. For this
denote by $\lambda_{[1]}=\frac{1}{\sqrt 2}
(\chi_{[1]}+\chi^*_{[1]})$ and $\lambda_{[3]}=-i\frac{1}{\sqrt
2}{}^* (\chi_{[1]} -\chi^*_{[1]})$. Notice that~$\lambda_{[1]}$
and~$\lambda_{[3]}$ are real. One has  ${}^{**}(\chi_{[1]}
-\chi^*_{[1]})=\chi_{[1]} -\chi^*_{[1]}$ and using also  the
def\/inition $d^\dag={}^*d^*$, equation \eqref{b1} writes
\begin{gather}
\label{b11} B_1\to B_1+ d\lambda_{[1]} + d^\dag\lambda_{[3]}.
\end{gather}
We now prove that for a $2$-form $B_1$ one can write a gauge
transformation as $B_1\to B_1 + d\lambda_{[1]}$, with $d^\dag
d\lambda_{[1]}=0$ but also as  $B_1\to B_1 +d^\dag\lambda_{[3]}$
with $d d^\dag \lambda_{[3]}=0$. Indeed, since $d^\dag
d\lambda_{[1]}=0$ this means that there exists a $3$-form
$\lambda_{[3]}$ such that $d\lambda_{[1]}=d^\dag\lambda_{[3]}$.
Thus, the gauge transformation can be written as $B_1\to B_1
+d^\dag\lambda_{[3]}$. Moreover, adding this two equivalent types
of gauge transformations one can write a ``general'' gauge
transformation for $B_1$ as in \eqref{b11}. This means that
\eqref{jauge-normale} and \eqref{g-trans} are compatible for the
$2$-forms even if the gauge parameter $\chi_{[1]}$ is complex.

We now show what are the constraints on the gauge parameters $k$,
$\tilde \chi$ and $\chi_m$ of the the gauge
transformations~\eqref{g-trans}. As before, the case of the scalar
and vector gauge parameter is simple, leading to
\[
\partial_m k = 0 \qquad \mbox{and}\qquad \Box \tilde \chi = 0.
\]
For the case of a $2$-form ${ B}^{(\pm)}$, one checks separately
the invariance under \eqref{g-trans-forme} of $(d{ B}^{(\pm)})^2$
and $(d^\dag { B}^{(\pm)})^2$ \footnote{Note that for the case of
a real $2$-form the gauge parameter $\chi'_{[1]}$ def\/ined in
\eqref{jauge-normale} is from the same reason subject to the
constraint  $d^\dag d \hat \chi_{[1]}=0$.}.
 This gives $d^\dag d \chi_{[1]}=0$.
Hence, these constraints on a general gauge parameter $\chi_{[p]}$
write
\begin{gather}
\label{gen} d^\dag d \chi_{[p]}= 0.
\end{gather}
In our particular case, using component notations, one has
\begin{gather} \label{g-fix}
\partial_m k = 0, \qquad \Box \tilde \chi = 0, \qquad \Box \chi_n - \partial_n  \partial_m \chi^m = 0.
\end{gather}

The strategy we adopt here is to f\/ind explicit forms of the
gauge parameters def\/ined in \eqref{g-trans} and subject to the
constraints \eqref{g-fix}.

{\it {\underline{II}}}. The second step of our analysis is to have
some transformations of a~multiplet into a~multiplet of same type,
transformation that can then be matched with the gauge
 transformation~(\ref{g-trans}). Since in Subsection \ref{deriv-mult}
 we have introduced the derivative multiplets (that transform like \3 multiplets), we can use them to do the job:
\begin{gather}
\label{pe-plac} {\boldsymbol \Xi}_{++}\to {\boldsymbol
\Xi}_{++}+{\cal D}\,{\boldsymbol \Xi}_{+-}.
\end{gather}
For instance,  one can use the derivative of a multiplet
${\boldsymbol \Xi}_{+-}=\big(\lambda_m, \tl, \tl_{mn}, \ttl_m
\big)$, writing thus~\eqref{pe-plac} as
\begin{gather}
 \varphi \to \varphi + \partial_m \lambda^m,  \qquad
 B_{m n} \to  B_{m n} + \partial_m \lambda_n - \partial_n \lambda_m
- i \varepsilon_{mnpq} \partial^p \lambda^q, \nonumber \\
 \tA_m \to \tA_m + \partial_m \tl + \partial^n \tl_{nm}, \qquad
 \ttP \to \ttP + \partial^m \ttl_m, \nonumber \\
  \ttB_{m n} \to \ttB_{m n} + \partial_m \ttl_n - \partial_n \ttl_m
- i \varepsilon_{mnpq} \partial^p \ttl^q.\label{g-trans1}
\end{gather}

{\it {\underline{III}}}. The last step of this programme is to
make (\ref{g-trans1}) a gauge transformation, that is to match it
with (\ref{g-trans}) and the conditions
(\ref{g-fix})\footnote{Since these equations have practically the
same form, it now becomes clear why we have chosen to work with
gauge transformations of type \eqref{g-trans} and not the gauge
transformations \eqref{jauge-normale} of the real f\/ields since
we work directly on the ${\boldsymbol \Xi}_{++}$.}.

First remark that the  actual matching of these transformations
implies a non-trivial condition for the parameters of the
transformation of the vector f\/ield, namely
\begin{gather} \label{g-conditions4}
\partial^n \tl_{nm} = \partial_m \chi
\end{gather}
and, since  $\tl_{nm}$ is antisymmetric, one has
\begin{gather} \label{g-conditions4pp}
\Box \chi =0.
\end{gather}

Now, imposing the conditions $(\ref{g-fix})$ on the set of gauge
parameters $\big(\lambda_m; \tl, \tl_{mn}; \ttl_m \big)$, one has
\begin{gather}
 \partial_m \; (\partial \cdot \lambda) = \partial_m \;
(\partial \cdot \ttl) = 0,  \label{g-conditions1}\\
\Box \lambda_m = \Box \ttl_m =0,
 \label{g-conditions2}\\
 \Box \tl + \partial^m \partial^n \tl_{nm} \equiv \Box \tl = 0.
\label{g-conditions3}
\end{gather}

   We have now to f\/ind explicit solutions of the gauge parameters which
   are compatible with all these constraints. In order to do this,
\begin{enumerate}\itemsep=0pt
\item[1)] we determine the solutions for $\tl, \chi$ satisfying
the constraints
 (\ref{g-conditions3}), (\ref{g-conditions4pp});
\item[2)] knowing $\chi$, we then construct   an anti-self-dual
$2$-form $\tilde \lambda_{mn}$ satisfying (\ref{g-conditions4});
\item[3)] we f\/inally f\/ind explicit solutions for  $\lambda_m,
\ttl_m$ satisfying (\ref{g-conditions1}), (\ref{g-conditions2}).
\end{enumerate}

Existence of these solutions  would prove at this step
compatibility between \3 and the Abelian gauge invariance.

$1.$ If  the scalar functions $\tl$ and $\chi$ depend only on the
space-time Lorentz  invariant $x_m x^m$, then the conditions
(\ref{g-conditions3}), (\ref{g-conditions4pp}) determine uniquely
their form, $\tl(x^2) \propto \chi(x^2) \propto 1/x^2$ up to some
additive constants. In the context of the symmetry
\eqref{algebra}, whose generators and transformation parameters
are $4$-vectors, it is somewhat natural to include dependence of a
$4$-vector $\xi_m$. Moreover, by analysing  solutions of
\eqref{g-conditions3} and \eqref{g-conditions4pp} when $\xi^2$ is
equal to or dif\/ferent from $0$, we f\/ind more general
conf\/igurations for $\lambda$ and $\chi$ when $\xi^2=0$ (see
\cite{io1}). Hence we explicitly treat this case in this
subsection.

We thus have to treat an equation of type
\begin{gather}
\label{ec-scalar} \Box f (x^2, x \cdot \xi)= 0,
\end{gather}
where $f$ denotes generically $\tilde \lambda$ or $\chi$. A
solution for this equation is given by
\begin{gather}
\label{sol-scalar} f (x^2, \; \xi \cdot x) = G(\xi \cdot x) + (\xi
\cdot x)^{-1} H\left(\frac{x^2}{(\xi \cdot x)}\right),
\end{gather}
where G and H are arbitrary functions.

This provides us an explicit form for the parameters $\tilde
\lambda$ or $\chi$
\begin{gather} \label{eqtl}
\tl(\xi \cdot x, x^2) = G_1(\xi \cdot x) + (\xi \cdot x)^{-1}
H_1\left(\frac{x^2}{(\xi \cdot x)}\right),
\\ \label{eqchi}
\chi(\xi \cdot x, x^2) = G_2(\xi \cdot x) + (\xi \cdot x)^{-1}
H_3\left(\frac{x^2}{(\xi \cdot x)}\right),
\end{gather}
completing thus step $1$ of the programme.

2. As already stated, we now have to f\/ind a form of $\tl_{mn}$
which satisf\/ies the constraint (\ref{g-conditions4}),
with~$\chi$ given by (\ref{eqchi}) above. A possible solution is
\begin{gather} \label{eqtlmn}
\tl_{m n}\big(\xi \cdot x, x^2\big) = x_{[m} \xi_{n]_-} F\big(\xi
\cdot x,  x^2\big),
\end{gather}
where the function $F$ is expressed in terms of $G_2$, $H_3$
appearing in (\ref{eqchi}):
\begin{gather}
F\big(\xi \cdot x, x^2\big) = - (\xi \cdot x)^{-2}
H_3\left(\frac{x^2}{(\xi \cdot x)}\right) + (\xi \cdot x)^{-1}
G_2(\xi \cdot x)
 -2(\xi \cdot x)^{-3} \int_0^{\xi \cdot x}\! G_2(t) t \, dt.
\end{gather}

3. Similarly to step $1$, we  now investigate possible solutions
for the gauge parameters $\lambda_m$,~$\ttl_m$, which satisf\/ies
equations \eqref{g-conditions1} and \eqref{g-conditions2}. We
consider them as functions of the vectors $x$ and~$\xi$ and, as
before, we assume $\xi^2=0$. Hence, the problem is reduced to
f\/inding explicit solutions for
\begin{gather}
\label{ec-vector}
\partial_m \partial_p {\cal A}^p (x,\xi)=0,\qquad
\Box {\cal A}_m (x,\xi)=0
\end{gather}
(which are just equations (\ref{g-conditions1}) and
(\ref{g-conditions2}), $\cal A$ standing for $\lambda_m$ or
${\tilde {\tilde \lambda}}_m$). Some solution of these equation is
given by
\begin{gather}
\label{sol-vector}
 {\cal A}_m (x, \xi)= g(\xi \cdot x) \xi_m + \alpha x_m +
\left(\frac{1}{(x^2)^2} \alpha_{m r} + \beta_{m r}\right) x^r +
\kappa \left(\frac{x^2}{(\xi \cdot x)^3} \xi_m - \frac{x_m}{(\xi
\cdot x)^2}\right),
\end{gather}
where $g$ is an arbitrary function, $\kappa$, $\alpha$, $\beta_{m
n}$ arbitrary constants and $\alpha_{m n}$ an arbitrary
anti-symmetric tensor.

Thus one can now write the following expressions for the last
gauge parameters $\lambda_m$ or ${\tilde {\tilde \lambda}}_m$
\begin{gather}
\lambda_m \big(\xi \cdot x,  x^2\big) = g_1(\xi \cdot x) \xi_m +
 \alpha x_m \nonumber\\
 \phantom{\lambda_m \big(\xi \cdot x,  x^2\big) =}{}+ \left(\frac{1}{(x^2)^2} \alpha_{m r} + \beta_{m r}\right) x^r
 +\kappa_1 \left(\frac{x^2}{(\xi \cdot x)^3} \xi_m - \frac{x_m}{(\xi \cdot x)^2}\right), \label{eqlambdam}
\\
\ttl_m \big(\xi \cdot x, x^2\big) = \tilde{\tilde g}_1(\xi \cdot
x)
 \xi_m  +
\tilde{\tilde \alpha} x_m \nonumber\\
\phantom{\ttl_m \big(\xi \cdot x, x^2\big) =}{}+
\left(\frac{1}{(x^2)^2} \tilde{\tilde \alpha}_{m r} +
 \tilde{\tilde \beta}_{m r}\right) x^r  + \tilde{\tilde \kappa}_1 \left(\frac{x^2}{(\xi \cdot x)^3} \xi_m -
\frac{x_m}{(\xi \cdot x)^2}\right).\label{eqttlm}
\end{gather}

  We have therefore obtained a proof of existence of gauge
  transformations which are compa\-tible with our symmetry.
  Nevertheless, we have found constrained forms of our gauge
  parameters~$\lambda_m$, $\tl$, $\tl_{mn}$ and resp.~$\ttl_m$ (equations \eqref{eqlambdam},
  \eqref{eqtl}, \eqref{eqtlmn} and resp.~\eqref{eqttlm}).

Before ending this section,
 let us mention that a discussion related to Noether
 currents was initiated in \cite{articol}, mostly in relation with
 to the case of algebra of conserved charges compared
 to the case where symmetries are grouped within Lie (super)algebras.
 Furthermore some conserved currents were explicitly calculated in~\cite{rausch}.

\section{Study of interaction possibilities}
\label{interactiuni}

In the previous section we have considered non-interacting terms
allowed by \3 invariance. Here we investigate the possibility of
interacting terms, terms which must have a degree in the f\/ields
higher than $2$ (thus not being possible to diagonalise them back
to kinetic terms). The main result is that for the bosonic
multiplets considered, no such interaction terms are allowed.

To approach this issue we make a systematic study of all
interaction possibilities for our multiplets. We f\/irst f\/ind
what are the f\/ields $\Psi$ (content and transformation laws)
that can couple to  multiplets in an invariant quadratic way. We
then express these f\/ields $\Psi$ as a function
$\Psi({\boldsymbol \Xi}_{++},{\boldsymbol \Xi}_{--},{\boldsymbol
\Xi}_{+-},{\boldsymbol \Xi}_{-+})$. We f\/ind this function to be
linear in the multiplets; hence the most general invariant terms
which can be constructed are quadratic and thus non-interacting.

\subsection{Possible couplings of a given multiplet}
\label{cuplaje-posibile}

We focus on the coupling of a ${\boldsymbol \Xi}_{++}$ multiplet,
the other cases being similar. Considering its f\/ield content,
the most general possibility of quadratic coupling with some set
of unknown f\/ields $\Psi$ is
\begin{gather}
\label{X++coup} {\cal L}({{\boldsymbol \Xi}_{++}, \Psi})= \varphi
\ttpsi + \ttP \psi + \frac14 B^{(+)mn} \ttpsi_{mn} +
  \frac14 \ttB^{(+)}_{mn} \psi^{mn} - \tA_m \tpsi^m
\end{gather}
with $\psi$, $\ttpsi$ two scalars, $\tpsi_m$ a vector and
$\psi_{mn}$, $\ttpsi_{mn}$ two $2$-forms which are self-dual. {\it
A priori} some of the f\/ields $\Psi$ can be set to zero.

To f\/ind the set of f\/ields $\Psi$ we impose that
(\ref{X++coup}) transforms as a total derivative. We f\/irst treat
the case where the f\/ields $\Psi$ contain no derivative terms.
Applying (\ref{transfo2})  directly,  one has the variation of
(\ref{X++coup}) and is able to prove (see \cite{io1}):
\begin{enumerate}\itemsep=0pt
\item[{\bf I:}] {\it If the $\psi$ f\/ields contain no derivative
terms and \eqref{X++coup} is invariant, then they  form a
multiplet of  type ${\boldsymbol \Xi}_{++}$.}
\end{enumerate}

Similar type of arguments lead to the same conclusion if the
$\psi$ f\/ields contain at most one derivative term.

Let us now allow a higher number of derivatives in the f\/ields
$\Psi$. For example, if one considers two derivatives, one can
write f\/ields of type
\[
 \psi= \Box \lambda,\ \psi_m=\alpha \Box \lambda_m+\beta \partial_m\partial_n
\lambda^n,\qquad \psi_{mn}=\alpha' \Box
\lambda_{mn}+\beta'\partial^p\partial_{[m}\lambda_{n]+p}.
\]
Generally speaking, if one consider an even number $n$ of partial
derivatives, the terms that can be added are
\[
 \psi = \Box^{\frac n2} \lambda,\qquad
  \psi_m= \alpha \Box^{\frac n2} \lambda_m + \beta \Box^{\frac{n-2}{2}}
  \partial^n \partial_m \lambda_n,\qquad \psi_{mr}=\alpha \Box^{\frac n2}
  \lambda_{mr}+ \beta \Box^{\frac{n-2}{2}} \partial^p\partial_{[m}\lambda_{n]+p}.
  \]
If $n$ is an odd number, one can construct
\[
\psi=\Box^{\frac{n-1}{2}} \partial^m \lambda_m,\qquad
 \psi_m=\alpha \Box^{\frac{n-1}{2}} \partial_m \lambda
 + \beta \Box^{\frac{n-1}{2}} \partial^p \lambda_{pm},\qquad
  \psi_{mp}= \Box^{\frac{n-1}{2}} \partial_{[m}\lambda_{n]+}.
 \]
Similar arguments lead to the same type of result {\bf I.}

We have thus seen which are the most general couplings of a given
multiplet. We  now have the set of f\/ields $\Psi$, its content
and transformation laws in terms of the $\lambda$ f\/ields. In the
remainder of this section we see what is the most general way one
can construct these $\Psi$ f\/ields out of the original
${\boldsymbol \Xi}_{\pm\pm}$ multiplets.

\subsection{Generalised tensor calculus}

Here also we focus on the ${\boldsymbol \Xi}_{++}$ multiplet, the
other cases being similar. So, what we look for is to express the
f\/ields $\psi$, $\psi_{mn}$, $\tpsi_m$, $\ttpsi$, $\ttpsi_{mn}$
of the previous subsection as functions of the f\/ields of the
multiplets ${\boldsymbol \Xi}_{++}$, ${\boldsymbol \Xi}_{--}$,
${\boldsymbol \Xi}_{+-}$, ${\boldsymbol \Xi}_{-+}$ and of their
derivatives.

After direct explicit calculations (see \cite{teza} for details)
one proves
\begin{enumerate}
\item[{\bf II.}] {\it The only function $\Psi$ with at most
f\/irst order derivatives in the f\/ields and  transforming as a
${\boldsymbol \Xi}_{++}$ multiplet is}
\begin{gather*}
\Psi ({\boldsymbol \Xi}_{++}, {\boldsymbol \Xi}_{+-}, {\boldsymbol
\Xi}_{+-},{\boldsymbol \Xi}_{--}) = \alpha {\boldsymbol \Xi}_{++}
+ \beta {\boldsymbol \Xi}_{--}^* + \gamma{\cal D} {\boldsymbol
\Xi}_{+-}+ \mu {\cal D} {\boldsymbol \Xi}_{-+}^*.
\end{gather*}
\end{enumerate}

Moreover, if one considers several copies of the same multiplet,
the conclusion does not change. The case of functions involving
higher number of derivatives does not change the f\/inal
conclusion either:
 the $\psi$ f\/ields can be obtained only linearly out
 of the four considered  bosonic multiplets. Comparing
 this result with the result of the previous subsection
 (which was stating that these $\psi$ f\/ields are the most
 general possibility to quadratically couple the multiplets
 to some arbitrary f\/ields) one concludes that no invariant terms
 of order higher than two in the f\/ields can be constructed. This means
 that {\it one cannot obtain invariant self-interacting terms for the
 bosonic multiplets ${\boldsymbol \Xi}_{\pm \pm}$}.

\section{Concluding remarks}
\label{4-perspective}

Even if we have proven in the previous section  impossibility of
writing invariant interaction terms, this  situation can be
compared  with  incompatibility for the usual electromagnetism,
where photons do not self-interact. Hence,  one may think of a
possibility of constructing  non-Abelian models.

More general possibilities of interaction have to be investigated
for a verdict on this issue. One might reconsider at this level
the fermionic multiplets of \cite{articol} and investigate a
possible interaction between them and  boson multiplets.
Furthermore, interactions with a dif\/ferent type of bosonic
multiplets (eventually more general multiplets) may be taken into
consideration. Let us also mention that considering  $p$-forms
with $p\ge 2$ implies
 a high rigidity for the interaction possibilities,
see \cite{henneaux}).

Some deeper analysis of possible mechanisms of elimination of
unphysical degrees of freedom of the f\/ields may be appropriate.
This may eventually involve some presence of ghosts, in connection
with suited quantif\/ication procedures.

In \cite{praga}, G.~Moultaka {\it et. al.} consider the algebra
\eqref{algebra} in arbitrary dimensions; this leads to the
implementation of a new, cubic symmetry at the level of $p$-forms.
Recalling that the $p$-forms of the bosonic multiplets couple
naturally to extended objects of dimension $(p-1)$
($(p-1)$-branes); one could seek for such an invariant theory for
interacting $p$-branes.

A dif\/ferent type of remark is to be made when one considers
dif\/ferent types of algebraic structure mentioned in
Section~\ref{sec1}: possible useful connections in between do not
seem very likely, since these algebraic structures are dif\/ferent
from the very def\/inition.

However, recalling the construction of \cite{plus4} of a
relativistic wave equation involving the cubic root of the
Klein--Gordon operator, let us mention that as Dirac equation is
related to SUSY, this cubic root of the Klein--Gordon operator is
related to fractional supersymmetry. As already mentioned here in
Subsection~\ref{conections}, in \cite{FSUSY5} for $1$-dimensional
fractional supersymmetry, the theory is localised leading to a
local fractional supergravity. In the case of the
$(3+1)$-dimensional model reviewed here, a generalisation in such
a direction constitutes a promising perspective of future work.

Finally, let us argue on the fact that maybe before  starting
 more f\/ield theoretical developments of this type of models,
it would be more important to have a closer look at their very
foundations, {\it i.e.}~the implications they would have  when
 compared to some of basic notions of f\/ield theory,
 like for example the assumptions of analicity
 of no-go theorems, canonical quantisation rules {\it etc}.
 This type of study may give stronger information of what may or
 may not be pertinent for coherent physical approaches.

\appendix

\section[Determination of the matrices $\sigma_{mnr}$]{Determination of the matrices $\boldsymbol{\sigma_{mnr}}$}

In this appendix we show how one obtains the connection
 \eqref{tanasa} between the generators $V$ and the SUSY supercharges $Q$;
 we thus obtain the explicit form of the matrices $\sigma_{mnr}$ appearing in equation~\eqref{tanasa}.

First recall equations \eqref{prima} and \eqref{adoua} of the
algebra \eqref{algebra} resp. SUSY algebra:
\begin{gather}
\left\{V_m, V_n, V_r \right \} = \eta_{m n} P_r +  \eta_{m r} P_n
+ \eta_{r n} P_m,\qquad \{ Q_\alpha, \bar Q_{\dot \beta} \} = 2
(\sigma^m)_{\alpha \dot \beta} P_m\label{amandoua}
\end{gather}
 (with $\alpha$, $\dot \beta =1,2$, see \cite{sohnius} for conventions on the SUSY algebra).

With use of the form of the Pauli matrices $\sigma^m$, the second
equation of \eqref{amandoua} writes explicitly
\begin{gather}
\left\{ Q_1, \bar Q_{\dot 1} \right\} = 2 (P_0 + P_3), \qquad
\left\{ Q_1, \bar Q_{\dot 2} \right\} = 2 (P_1 - i P_2), \nonumber\\
\left\{ Q_2, \bar Q_{\dot 1} \right\} = 2 (P_1 + i P_2), \qquad
\left\{ Q_2, \bar Q_{\dot 2} \right\} = 2 (P_0 -
P_3).\label{desfacut}
\end{gather}
These equations allow one to express the momentums $P_m$ as a
function of the SUSY algebra anticommutators:
\begin{gather}
P_0  =  \frac14 \big( \{ Q_1, \bar Q_{\dot 1}\} +  \{ Q_2, \bar
Q_{\dot 1}\}\big),\qquad
P_1 =\frac14 \big( \{ Q_1, \bar Q_{\dot 2}\} +  \{ Q_2, \bar Q_{\dot 1}\}\big),\nonumber\\
P_2  =  \frac i4 \big( \{ Q_1, \bar Q_{\dot 2}\} - \{ Q_2, \bar
Q_{\dot 1}\}\big),\qquad P_3 =  \frac14 \big( \{ Q_1, \bar Q_{\dot
1}\} -  \{ Q_2, \bar Q_{\dot 1}\}\big).\label{solutia}
\end{gather}
One can now insert equations \eqref{solutia} in the f\/irst
equation of \eqref{amandoua}. For example, one has
\begin{gather*}
\{ V_0, V_0, V_0 \}= 3 P_0
\end{gather*}
and, inserting the f\/irst of equations \eqref{solutia} one has
\begin{gather*}
(\sigma_{000}){}^{1,\dot 1}= \frac 34, (\sigma_{000}){}^{2,\dot
2}= \frac 34.
\end{gather*}
Similarly one gets the rest of the entries of the matrices
$\sigma_{mnr}$. Thus the non-zero entries of these matrices are:
\begin{alignat}{3}
& (\sigma_{000}){}^{1,\dot 1}= \frac 34, && (\sigma_{000}){}^{2,\dot 2}= \frac 34, &\nonumber\\
& (\sigma_{001}){}^{1,\dot 2}= \frac 14, && (\sigma_{001}){}^{2,\dot 1}= - \frac 14, &\nonumber\\
&(\sigma_{002}){}^{1,\dot 2}= \frac i4, && (\sigma_{001}){}^{2,\dot 1}= - \frac i4, &\nonumber\\
& (\sigma_{003}){}^{1,\dot 1}= \frac 14, && (\sigma_{003}){}^{2,\dot 2}= -\frac 14, &\nonumber\\
& (\sigma_{111}){}^{1,\dot 2}= -\frac 34, && (\sigma_{111}){}^{2,\dot 1}= - \frac 34, &\nonumber\\
& (\sigma_{112}){}^{1,\dot 2}= -\frac i4, && (\sigma_{112}){}^{2,\dot 1}=  \frac i4, &\nonumber\\
& (\sigma_{113}){}^{1,\dot 1}= -\frac 14, && (\sigma_{003}){}^{2,\dot 2}= \frac 14, &\nonumber\\
& (\sigma_{222}){}^{1,\dot 2}= - \frac{3i}{4}, && (\sigma_{222}){}^{2,\dot 1}= \frac{3i}{4}, &\nonumber\\
& (\sigma_{223}){}^{1,\dot 1}= -\frac 14, && (\sigma_{223}){}^{2,\dot 2}= \frac 14, &\nonumber\\
& (\sigma_{333}){}^{1,\dot 1}= -\frac 34, && (\sigma_{333}){}^{2,\dot 2}= \frac 34, &\nonumber\\
& (\sigma_{011}){}^{1,\dot 1}= -\frac 14, && (\sigma_{011}){}^{2,\dot 2}= -\frac 14, &\nonumber\\
& (\sigma_{022}){}^{1,\dot 1}= -\frac 14, && (\sigma_{022}){}^{2,\dot 2}= -\frac 14, &\nonumber\\
& (\sigma_{033}){}^{1,\dot 1}= -\frac 14, && (\sigma_{033}){}^{2,\dot 2}= -\frac 14, &\nonumber\\
& (\sigma_{033}){}^{1,\dot 1}= -\frac 14, && (\sigma_{033}){}^{2,\dot 2}= -\frac 14, &\nonumber\\
& (\sigma_{122}){}^{1,\dot 2}= -\frac 14, && (\sigma_{122}){}^{2,\dot 1}= - \frac 14, &\nonumber\\
& (\sigma_{133}){}^{1,\dot 2}= - \frac{1}{4}, && (\sigma_{133}){}^{2,\dot 1}=- \frac{1}{4}, &\nonumber\\
& (\sigma_{233}){}^{1,\dot 2}= - \frac{i}{4},\qquad &&
(\sigma_{233}){}^{2,\dot 1}=- \frac{i}{4}. &\label{matricele}
\end{alignat}

\subsection*{Acknowledgements}

I would like to acknowledge G.~Moultaka and M.~Rausch de
Traubenberg for their important help. I would also like to thank
R.~Kerner for very useful remarks.

\LastPageEnding

\end{document}